\documentclass[fp,singlecolumn]{jpsj3}
\usepackage{txfonts}

\title{Matter-Wave Fields for Double-Slit Atom Interferometry: Variational Versus Exact Solitons}

\author{Isaiah Ndifon Ngek$^1$, Alain Mo\"ise Dikand\'e$^1$\thanks{dikande.alain@ubuea.cm}, and Alain Brice Moubissi$^2$}
\inst{$^1$Laboratory of Research on Advanced Materials and Nonlinear Sciences (LaRAMaNS),
Department of Physics, Faculty of Science, University of Buea, P.O. Box 63, Buea, Cameroon \\
$^2$D\'epartement de Physique, Facult\'e des Sciences, Universit\'e 
des Sciences et Techniques de Masuku (USTM) BP. 943 Franceville, Gabon} %\\

\abst{A major challenge in the theoretical modeling of double-slit interferometry involving matter-wave fields is the appropriate waveform to be assigned to this field. While all the studies carried out to date on this issue deal with variational fields, experiments suggest that the optical field is generated by splitting a single-hump Bose\text{--}Einstein condensate into two spatially and temporally entangled pulses indicating the possibility of fully controlling the subsequent motion of the two output pulses. To probe the consistency of variational and exact soliton solutions to the field equation, we solve the Gross\text{--}Pitaevskii equation with an optical potential barrier assumed to act as a beam splitter, while including gravity. The exact solution is compared with the two most common variational wavefunctions, namely, the Hermite\text{--}Gaussian and super-sech modes. From numerical simulations, evidence is given of the exact solution as being the most appropriate matter-wave structure that provides a coherent description of the generation and spatio-temporal evolution of matter-wave optical fields in a hypothetical implementation of double-slit atom interferometry.}

\begin{document}
\maketitle

\section{Introduction}

Bound-soliton pairs provide interesting transmission channels for multiplexed high-intensity pulse trains~\cite{tang1,tang2,tang3,tang4,grelu1,dika1}. Among them, bisolitons have attracted a great deal of attention following their prediction~\cite{kumar1,Pare99,Maruta01,Maruta02,malomed97,Russell04,Bao04,Kockaert05} and observation~\cite{Statmann05,chong08} in several distinct optical media. In optical fibers, for instance, such structures originate from splitting~\cite{infeld1,infeld2} a single-pulse optical soliton using a quadratic chirp, which leads to a soliton molecule whose intensity profile exhibits two temporal and spectral peaks with a finite phase difference between them~\cite{Pare99,chong08}. Similar objects have been predicted and observed in Bose\text{--}Einstein condensate (BEC) systems~\cite{fuji1,shin04,Sin05}, where they result from a beam splitter acting on a bright matter-wave pulse, which creates two co-propagating matter-wave pulses. \\ 
Bisolitons are of fundamental importance in BECs since they provide appropriate wave structures needed for the experimental achievement of an atomic equivalent of double-slit holography~\cite{pezze05,pezze06}. In such an experiment, we can imagine two spatially and temporally entangled matter-wave pulses generated from the splitting of a single condensate into two atomic populations, namely in nearly degenerate hyperfine levels (see for example ref. \cite{stamper98}). Generally a far-off resonant laser barrier provides an ideal scheme for such splitting~\cite{pezze05,pezze06}. In theory, this most often has been described by considering a single matter-wave pulse passing through a potential barrier created by a laser field, with only a fraction of the macroscopic atomic system allowed to cross the barrier while the other fraction remains trapped on the other side of the barrier. \\
A pioneering model for such process was proposed in ref.~\cite{raghavan99}, in terms of a single macroscopic atomic population moving in a double-well optical field of a finite barrier. In this model, the system dynamics leads to two unbalanced population fractions from a single initial atomic population trapped in the two wells surrounding the potential barrier, with one fraction in the left well and the other fraction in the right well, after tunneling through the finite barrier. In the approach of ref.~\cite{raghavan99}, however, the focus was on the quantum tunneling dynamics of atoms, particularly the associated Josephson effect related to the phase difference between the wavefunctions of the two condensates. Nevertheless, having at hand an analytical expression for the full condensate wavefunction can also be useful for experiments. This latter problem can be addressed by considering the Gross\text{--}Pitaevskii (GP) equation with an antiharmonic (i.e., a repulsive harmonic) optical potential representing the laser barrier, which is actually equivalent to the nonlinear Schr\"odinger equation (NLSE) describing the propagation of pulse signals in an optical fiber with a quadratic chirp~\cite{Pare99,Maruta01,Maruta02,Russell04,Bao04,Kockaert05,turits1,kaup1}. \\
In recent works dealing with matter-wave solitons within the framework of the GP equation with a harmonic optical potential, emphasis has mostly been on variational considerations, where the macroscopic condensate wavefunction is approximated by the Hermite\text{--}Gaussian mode or a super-sech pulse~\cite{zang1,konotop1,gardiner1}. Still, the variational treatment is based on perturbation theory, both in the choice of the trial solution (despite its localized pulse shape, the Hermite\text{--}Gaussian mode is not a solution to the GP equation) and in the formulation of the time evolution of the variational soliton's collective coordinates~\cite{malomed2}. \\
In this work, we address the issue by considering two distinct variational solutions, namely, the super-sech and Hermite\text{--}Gaussian modes~\cite{fuji1,zang1,konotop1,gardiner1} on one hand, and the exact one-soliton solution to the associated GP equation obtained by a non-isospectral inverse-scattering transform (NIST) method~\cite{dikande1} on the other hand. A gravitational potential is included with the aim of taking into consideration the possible relative acceleration of one pulse in the post-created bisoliton with respect to the other due to the free-fall motion of atoms in the condensate fraction exposed to gravity~\cite{wadati99,wadati01,coq01,robins05}. In effect, because of the finite mass of atoms, gravity turns out to be an additional fundamental ingredient in matter-wave interferometry as previously emphasized (see for example refs.~\cite{pezze05,pezze06}). \\

\section{Exact One-Soliton Solution}
Consider a system of $N$ weakly interacting identical atoms of mass $m$. Because of their interactions, the $N$ atoms form an ensemble falling freely under gravity but also experiencing a repulsive force due to a potential barrier of finite height. In the mean-field picture, the quantum states of such a macroscopic system of bosons can be represented by a wavefunction $A(z,\tau)$, where $z$ represents the direction of the free fall of atoms and $\tau$ is the time variable. The GP equation governing the spatial and temporal evolution of $A(z,\tau)$ is given by
\begin{equation}
i\hbar A_{\tau}+\frac{\hbar^2}{2m}A_{zz}+2R|A|^2A-V(z) A=0, \label{eq1}
\end{equation}
where $R$ is the mean\text{--}field interatomic interaction coefficient related to the s-wave scattering length. $V(z)$ is the total external potential for the nonlinear Schr\"odinger-type Eq. (\ref{eq1}), it is given as the sum of the repulsive harmonic potential $V_{op} =-m\omega^2_g\,z^2$ and the linear potential $mg\,z$ accounting for the gravitational field
\begin{equation}
 V(z)=mgz-m\omega^2_g\,z^2, \label{eq2}
\end{equation}
with $g$ the acceleration of gravity and $\omega_g$ a characteristic frequency. To obtain the exact solution to Eq.~(\ref{eq1}) using the NIST method, it is useful to introduce the following dimensionless variables  
\begin{eqnarray}
T &=& \frac{R}{\hbar}\tau,  \hskip 0.4truecm x= \sqrt{\frac{2Rm}{\hbar^2}}\,z, \label{eq3a} \\
\tilde{V}(x)&=& -\frac{\mu_2 \hbar^2}{8 R^2m}\,x^2 + 
\frac{\mu_1}{R}\sqrt{\frac{\hbar^2}{2Rm}} \,x, \label{eq3b} \\
\mu_1  &=& mg,\hskip 0.4truecm \mu_2  = 4m\omega_g^2. \label{eq3c}
\end{eqnarray}
With these new variables, the GP equation in Eq.(\ref{eq1}) becomes
\begin{equation}
i A_T + A_{xx} - 2\left[\tilde{V}(x) - \vert A \vert^2\right] A =0, \label{eq4}
\end{equation}
which is simply the self-focusing NLSE with a nonlocal (i.e., spatially varying) external potential $\tilde{V}(x)$. When $\mu_2 =0$ and $\mu_1\neq0$, Eq.~(\ref{eq4}) describes the macroscopic dynamics of a free-falling BEC. The same equation was proposed to model the dynamics of plasmas in a gravitational field, and its exact one and n-soliton solutions have been obtained~\cite{liu1} by means of the Inverse-Scattering Transform (IST). On the other hand, when $\mu_1= 0$ but $\mu_2\neq0$, Eq.~(\ref{eq4}) becomes the NLSE with an external space-varying repulsive harmonic potential. Exact soliton solutions for this latter physical context have also been obtained, using an improved IST technique that overcomes the non-isospectral character of the associated Lax-pair-type eigenvalue problem~\cite{bala1}. \\
When both $\mu_2$ and $\mu_1$ are nonzero, the perturbed NLSE Eq.~(\ref{eq4}) gives rise to an interesting problem in mathematical physics whose solution requires a subtle combination of the methods proposed independently for the two particular cases: $\mu_2\neq 0$ but $\mu_1=0$~\cite{bala1,nakk1,nak2,nak3}, and $\mu_2=0$ with $\mu_1\neq 0$~\cite{liu1}. In the general picture of the IST~\cite{ablo1,dodd} the first step is the construction of two pairs of coupled linear eigenvalue equations associated with the perturbed NLSE Eq.~(\ref{eq4}). Thus, let us consider the two following coupled pairs of eigenvalue problems~\cite{dikande1}  
\begin{eqnarray}
u_x + i\,\lambda\,u - A(x, T)\,v &=&0 , \nonumber \\ 
v_x - i\,\lambda\,v + A^{\star}(x, T)\,u &=&0 , \label{eq5a}
\end{eqnarray}
\begin{eqnarray}
u_T - P\,u - Q\,v &=& 0, \nonumber \\
v_T - W\,u + P\,v&=&0,  \label{eq5b}
\end{eqnarray}
where $u(x, T)$ and $v(x, T)$ are space-time-dependent functions that form a two-component spinor with $\lambda$ the associate eigenvalue, while
$P(x,T)$, $Q(x,T)$ and $W(x,T)$ are unknown functions that can be computed from the condition of compatibility of the coupled pairs of equations. 
The quantity $A(x, T)$, which is precisely the field function of our main interest, is here regarded as a scattering potential for the Lax-pair type
eigenvalue problem. In the spirit of the IST~\cite{ablo1,dodd}, exact solutions to the eigenvalue problems in Eqs.~(\ref{eq5a}) and (\ref{eq5b}) follow from the scattering matrix for a specific choice of the scattering potential. \\ 
For a localized (i.e., pulse-shaped) solution, we will require that the scattering potential vanishes asymptotically as $\vert x 
\vert \rightarrow \infty$ at any propagation time. Such
asymptotic behavior is only possible for a judicious choice~\cite{sats} of the initial 
scattering potential $A(x, T=0)$, and from this choice the time evolution of the scattering data can easily be computed 
via Gelfand-Levitan-Marchenko~\cite{ablo1,dodd} integral equations. Note that in this Fourier-type integral 
equation, the scattering matrix stands for a propagator governing the time evolution of the initial space-dependent bright matter-wave structure.\\ 
To determine conservation laws for the spectral set associated with the inverse scattering problem, we need appropriate compatibility conditions 
between equations of the eigensystem Eqs.~(\ref{eq5a}) and (\ref{eq5b}). For this purpose, we differentiate Eq.~(\ref{eq5a}) with respect to $T$ and Eq.~(\ref{eq5b}) with respect 
to $x$, and compare the resulting coupled sets. This yields the following secular equation for the spectral parameter $\lambda$ 
\begin{equation}
\tilde{V}_x - \lambda_T  - 2(i\lambda)^2_x=0. \label{eq6}
\end{equation}
It follows that, unlike the homogeneous NLSE for which the IST involves a spectral parameter that is constant in time and space~\cite{ablo1,dodd}, the compatibility equation in Eq.~(\ref{eq6}) suggests a non-isospectral scattering problem for Eq.~(\ref{eq4}). An important consequence of the non-universality of conservation laws is the possibility of multiple solution to the secular Eq.(\ref{eq6}), namely we can peak a solution by the method of separation of variables which amounts to introduce two distinct functions, one depending on space variable and the other on time variable, i.e.
\begin{equation}
\lambda(x, T)= e(x)\,f(T). \label{eq7}
\end{equation}    
In Eq.~(\ref{eq6}), the latter solution gives rise to the following space-time evolution equation for the two factors of the spectral parameter
\begin{equation}
f_T = \frac{\tilde{V}_x}{e} - \frac{2(f^2)}{e} (e^2)_x . \label{eq8}
\end{equation}
It is useful at this step to stress that if $f(T)$ was not time-dependent,
it would have appeared as a constant factor in Eq.~(\ref{eq7}) and so 
the eigenvalue $\lambda$ would be time-independent but varying with $x$, in
accordance with Eq.~(\ref{eq8}). Next, one can verify from Eq.~(\ref{eq7}) that 
allowing $\lambda$ to change only in time implies that $e_x=0$ thus the 
dependence of the external potential $\tilde{V}(x)$ on the space variable $x$ should be linear. In this case, Eq.~(\ref{eq8}) implies that $\lambda_T\equiv constant$, which is the result of the IST for the linear space-dependent external potential studied in ref.~\cite{liu1}. \\
To find a solution to Eq.~(\ref{eq6}) in the general case of an external potental wih a linear plus quadratic term, we impose the following constraints  
\begin{equation}
(e^2)_x= -\alpha\,e, \hskip 0.4truecm \tilde{V}_x= \beta\,e, \label{eq9}
\end{equation}
where $\alpha$ and $\beta$ are two arbitrary real constants. With these constraints Eq.~(\ref{eq6}) reduces to
\begin{equation}
f_T= -2\alpha\,f^2 + \beta. \label{eq10}
\end{equation}
Hence, the solutions to Eq.~(\ref{eq8}) are
\begin{eqnarray}
e(x) &=& -\frac{1}{2}\alpha\, x + \alpha_0, \nonumber \\
\tilde{V}(x)&=& \left(-\alpha\,x^2 + 4\,\alpha_0\,x\right)\frac{\beta}{4} + 
\beta_0,  \label{eq11}
\end{eqnarray} 
where $\alpha_0$ and $\beta_0$ are two integration constants. Given that $\tilde{V}(x)$ 
is explicitly defined in~(\ref{eq2}), equating coefficients of similar terms in this function and in~(\ref{eq11}) leads to
\begin{equation}
\alpha= \frac{\mu_2 \hbar^2}{4R^2m\beta}, \hskip 0.4truecm\alpha_0= \frac{4\mu_1}{R\beta}\sqrt{\frac{\hbar^2}{2mR}}, \hskip 0.4truecm \beta_0= 
0, 
\label{eq12}
\end{equation}
where $\alpha$ and $\beta$ are two integration constants. Actually the explicit form for $\tilde{V}(x)$ imposes a value on $\alpha$, while we are allowed to set $\beta=1$ and in this way all coefficients in the spatial component of $\lambda$ are unambiguously determined. For the temporal component, we integrate~(\ref{eq10}) to 
obtain
\begin{equation}  
f(T)= \frac{f(0) + f_0\,\tanh(\kappa\,T)}{f_0 + f(0)\,\tanh(\kappa\,T)}\,f_0, 
\label{eq13a}
\end{equation}
where 
\begin{equation}
f_0=\frac{2R\sqrt{m}}{\sqrt{2\hbar^2 \mu_2}}, \hskip 0.4truecm \kappa=1/f_o, \label{eq13b}
\end{equation}
with $f(0)$ the initial value determined by the scattering potential of the 
NIST. \\
To reconstruct the time evolution of the scattering data, we simply follow the standard considerations of the IST~\cite{dodd}, namely, we define a 
scattering matrix $G_s$ whose spectral problem involves a finite number ($N_b$) of bound states and a continuum; 
\begin{eqnarray}
&G_s&\left[x_1, T \right]=
i\sum_{\ell}^{N_b} {\frac{J_2[\lambda_{\ell}(x_1, T)]}{J_1'[\lambda_{\ell}(x_1, 
T)]}\,\exp [i\, T_1\lambda_{\ell}(x_1, T)]} \nonumber \\
&+& \frac{1}{2\pi}\int_{-\infty}^{\infty}{\frac{J_2[\lambda(x_1, 
T)]}{J_1[\lambda(x_1, T)]}\,\exp[i\, T_1\lambda(x_1, 
T)]\,d[\lambda(x_1, T)]}, \nonumber \\
\label{eq14}
\end{eqnarray}
\begin{equation}
x_1 = \int_{x_0}^x{e(x')dx'},
\label{eq14a}
\end{equation}
where the $J_i$ are the amplitudes of the Jost functions~\cite{bala1} which are the solutions to the non-isospectral eigenvalue problems Eqs.~(\ref{eq5a}) and (\ref{eq5b}), and the indices $\ell$ refer to the discrete eigenstates of the scattering matrix. With the help of the propagator Eq.~(\ref{eq14}), we can formulate the inverse transform as follows in terms of the Gelfand-Levitan-Marchenko integral equations~\cite{dodd}
\begin{eqnarray}
K_1(x_1, x_1',T))&=& G^{\star}_s(x_1 + x_1',T) \nonumber \\
                  &-& \int_{x_1}^{\infty}{K_2^{\star}(x_1, x_1'', T)G_s^{\star}(x_1'+ X_1'',T)dx_1''}, 
\nonumber \\
K_2(x_1, x_1', T)&=&- \int_{x_1}^{\infty}{K_1(x_1, x_1'', T)G_s(x_1'+ x_1'', 
T)dx_1''}. \nonumber \\
\label{eq15}
\end{eqnarray}
These two integral equations are solved exactly once an explicit shape 
for the scattering potential $A(x_1, 0)$ is chosen, leading to the following general solution for the 
nonlinear evolution equation~(\ref{eq4}) 
\begin{equation}
A(x, T)=2e(x)\,K_1\left[
\int_{x_0}^x{e(x')dx'},\int_{x_0}^x{e(x')dx'},T\right], \label{eq16a}
\end{equation}
with $x_0$ an arbitrary initial position of the field $A(x, T)$. \\ 
Given that our interest are bright matter-wave structures, it is legitimate to choose an initial solution with a pulse shape. Following Satsuma and Yajima~\cite{sats}, such an appropriate initial solution with a pulse shape is given exactly by
\begin{equation}
A(x, T=0)= -2e(x)\,sech\,\left(2 \int_{x_0}^x{e(x')dx'}\right). \label{eq16b}
\end{equation}
In quantum mechanics, this sech function represents a reflectionless potential and hence its boundstate spectrum possesses a twofold 
degenerate mode with the eigenvalues 
\begin{equation}
[f(0), f^{\star}(0)]=[i, -i], \label{eq17}
\end{equation} 
the structure of which suggests that the time-dependent part $f(T)$ of the non-isospectral eigenvalue of the NIST must be a complex function, namely 
\begin{equation}
f(T)= f_{re}(T) + if_{im}(T). \label{eq18}
\end{equation}
According to Eq.~(\ref{eq17}), Eq.~(\ref{eq18}), and the general expression for $f(T)$ 
given by Eqs.~(\ref{eq13a})-(\ref{eq13b}), the initial values of the real and imaginary 
parts of $f(T)$ should be
\begin{equation}
f_{re}(0)= 0, \hskip 0.4truecm f_{im}(0)=1. \label{eq19}
\end{equation}
Therefore, from~(\ref{eq18}) we are allowed to write
 \begin{eqnarray}
 f_{re}(T)&=& \frac{1}{\eta}\,\frac{(1 + \eta^2)\tanh(\kappa T)}{1 + 
\eta^2\tanh^2(\kappa T)}, \nonumber \\
 f_{im}(T)&=& \frac{sech^2(\kappa T)}{1 + \eta^2\tanh^2(\kappa T)}, \label{eq20}
 \end{eqnarray}
 \begin{equation}
 \eta= 1/\kappa. \label{eq21}
 \end{equation}
Also, the scattering potential~(\ref{eq16b}) permits a complete formulation of the
eigenstates of the NIST, including its eigenvalue spectrum and the set of 
associated eigenfunctions. With this scattering matrix, we construct the 
time-dependent solution to our inhomogeneous NLSE~(\ref{eq4}) from the integral 
equation~(\ref{eq16a}). We find
 \begin{equation}
 A(x, T)=A_0(x, T)sech\,2\Phi(x, T)\,\exp[-i\varphi(x, T)], \label{eq22}
 \end{equation}
 with:
 \begin{equation}
 A_0(x, T)= \vartheta_0\,e(x)f_{im}(T),  \label{eq23}
 \end{equation}
 \begin{eqnarray}
 \Phi(x, T)&=& - f_{im}(T)\int_{x_0}^x{e(x')dx'} + \ln\sqrt{\frac{\vert 
\vartheta_0 \vert}{2f_{im}(T)}} \nonumber \\
               &+& 
\frac{2\mu_1}{R}\sqrt{\frac{2\hbar^2}{mR}}\int_0^T{\Im{[f^2(T')]}dT'}, \label{eq24}
\end{eqnarray}
\begin{eqnarray}	       
 \varphi(x, T)&=&  2f_{re}(T)\int_{x_0}^T{e(x')dx'}  
\nonumber \\
               &+& 
\frac{4\mu_1}{R}\sqrt{\frac{2\hbar^2}{mR}}\int_0^T{\Re{[f^2(T')]}dT'}, \label{eq25}
 \end{eqnarray}   
 where $\vartheta_0$ in~(\ref{eq23}) is a constant amplitude that follows from 
the explicit form of the scattering matrix via the Gelfand-Levitan-Marchenko integral equations. The exact expression 
for this parameter resulting from the above treatment is  
 \begin{equation}
 \vartheta_0= \sqrt{\frac{\omega_o}{\omega_o^{\star}}}, \hskip 0.4truecm 
\omega_o=\frac{J_2(0)}{J_1'(0)}=const., \label{eq26}
 \end{equation}
 where $\omega_o/\omega_o^{\star}$ is the ratio of two arbitrary constant 
amplitudes of the Jost functions.  

\section{Variational Versus Exact Matter-Wave Solitons}
\begin{figure*}\centering
\begin{minipage}[b]{0.5\linewidth}
\includegraphics[width=\textwidth]{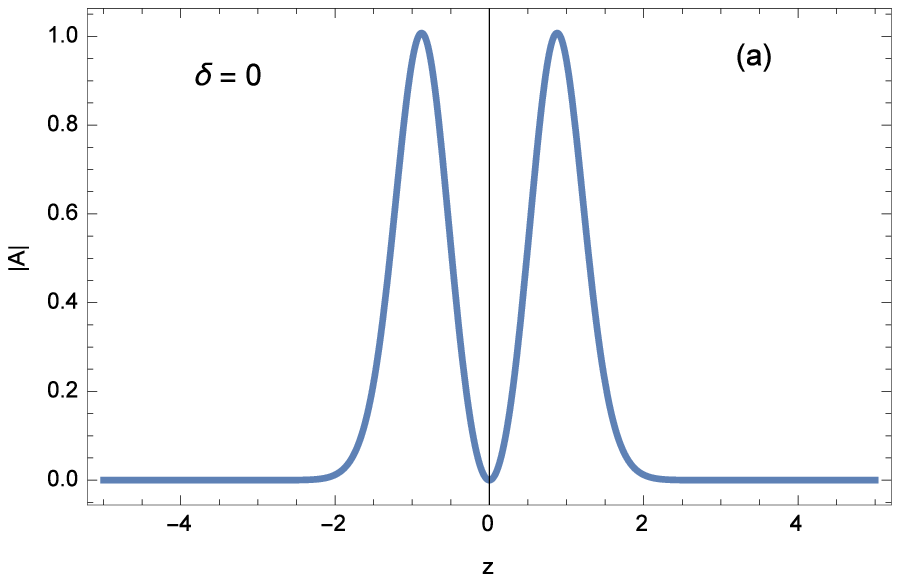}
\end{minipage}%
\begin{minipage}[b]{0.5\linewidth}
\includegraphics[width=\textwidth]{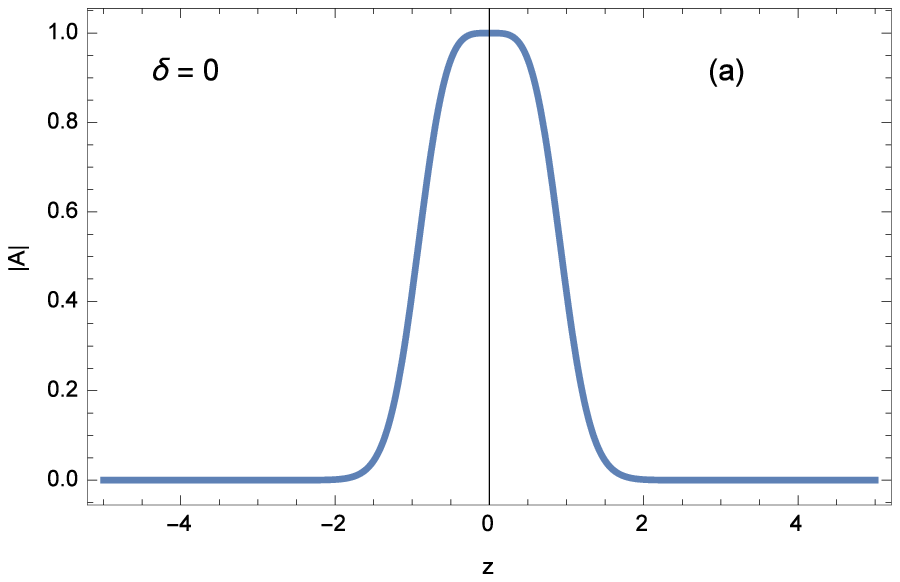}
\end{minipage}

\begin{minipage}[b]{0.5\linewidth}
\includegraphics[width=\textwidth]{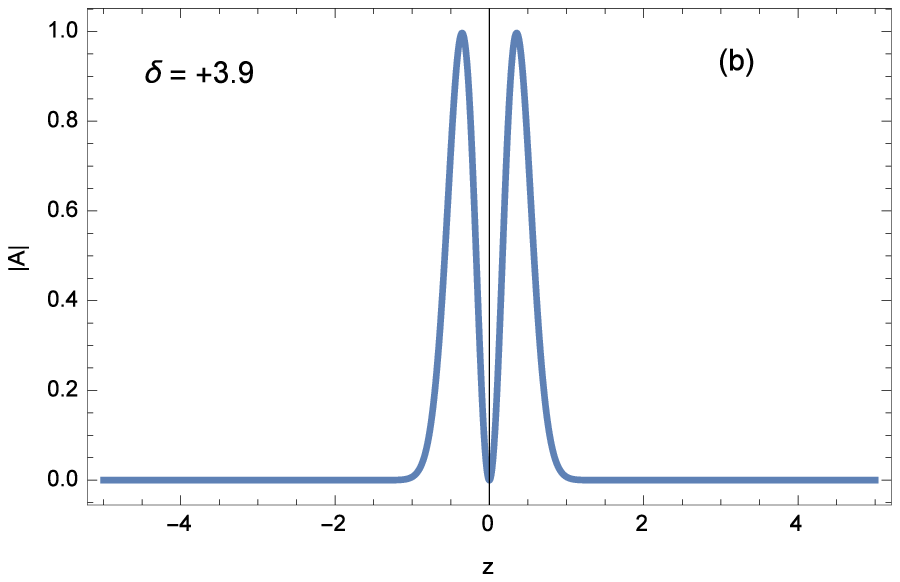}
\end{minipage}%
\begin{minipage}[b]{0.5\linewidth}
\includegraphics[width=\textwidth]{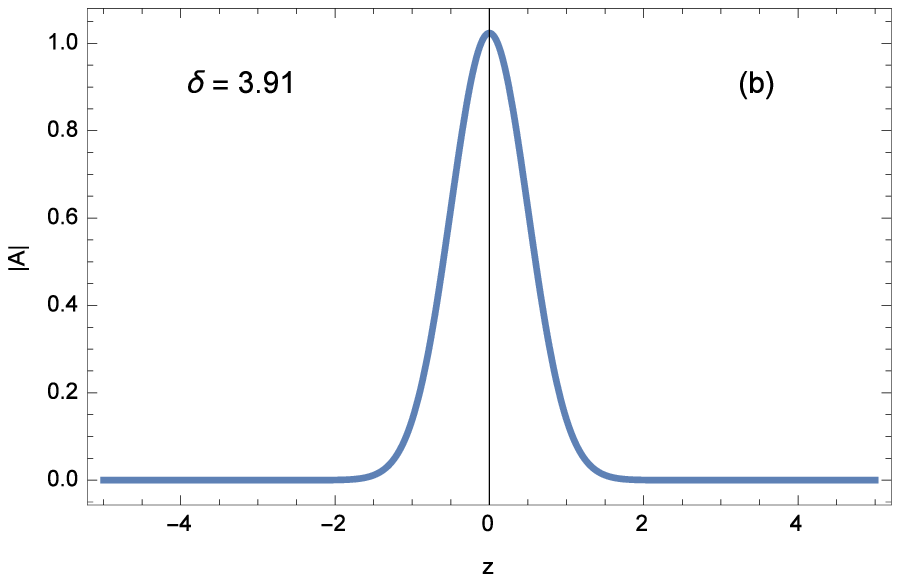}
\end{minipage}

\begin{minipage}[b]{0.5\linewidth}
\includegraphics[width=\textwidth]{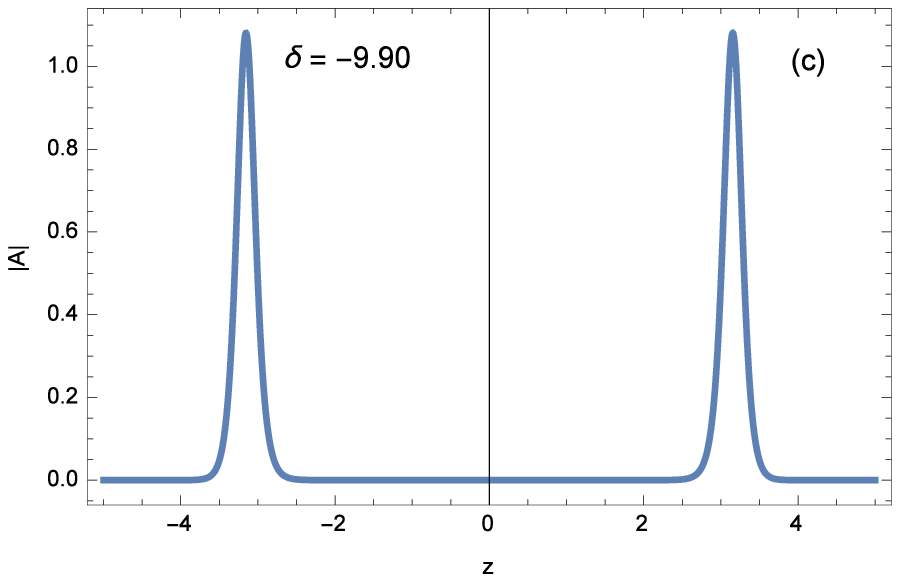}
\end{minipage}%
\begin{minipage}[b]{0.5\linewidth}
\includegraphics[width=\textwidth]{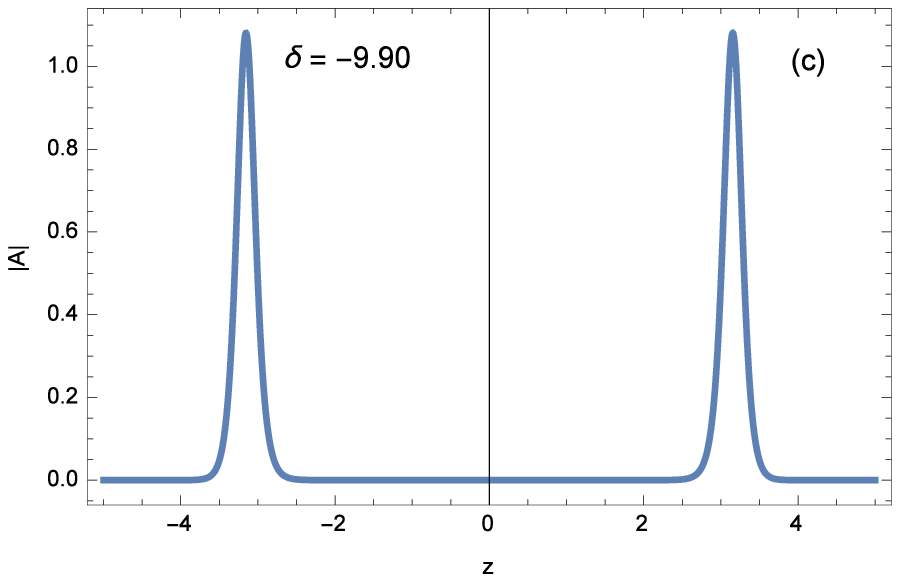}
\end{minipage}

\caption{\label{fig:one} (Color online) Spatial profiles of the exact bisoliton intensity (left) and of the equivalent "super-sech" ansatz (right) for $\mu_1=0$, $\mu_2=0.2$, and three distinct values of $\delta$.}
\end{figure*}

\begin{figure*}\centering
\begin{minipage}[b]{0.5\linewidth}
\includegraphics[width=\textwidth]{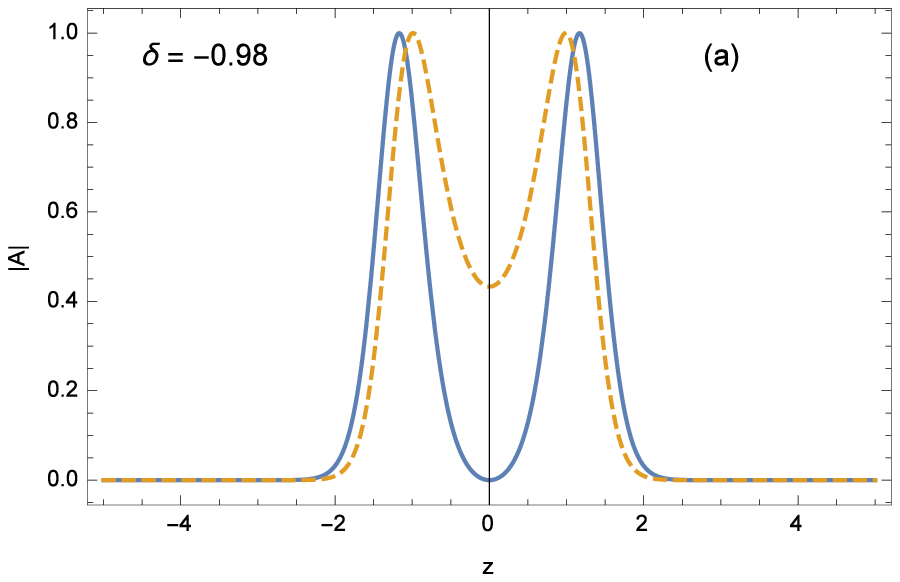}
\end{minipage}%
\begin{minipage}[b]{0.5\linewidth}
\includegraphics[width=\textwidth]{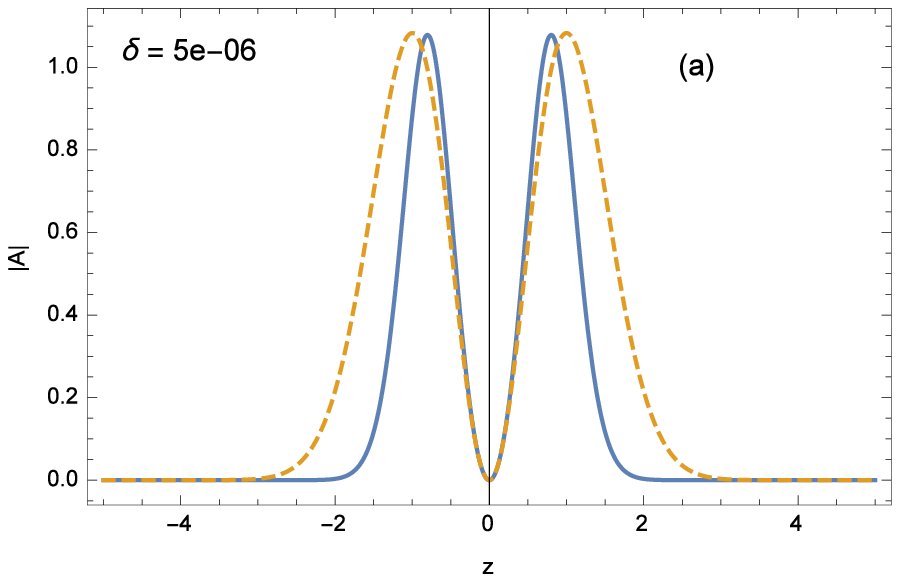}
\end{minipage}

\begin{minipage}[b]{0.5\linewidth}
\includegraphics[width=\textwidth]{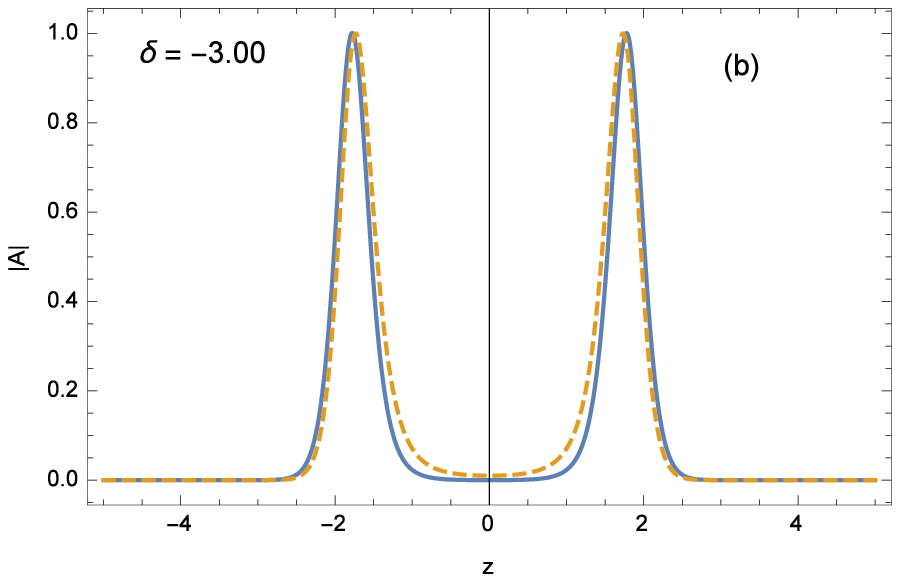}
\end{minipage}%
\begin{minipage}[b]{0.5\linewidth}
\includegraphics[width=\textwidth]{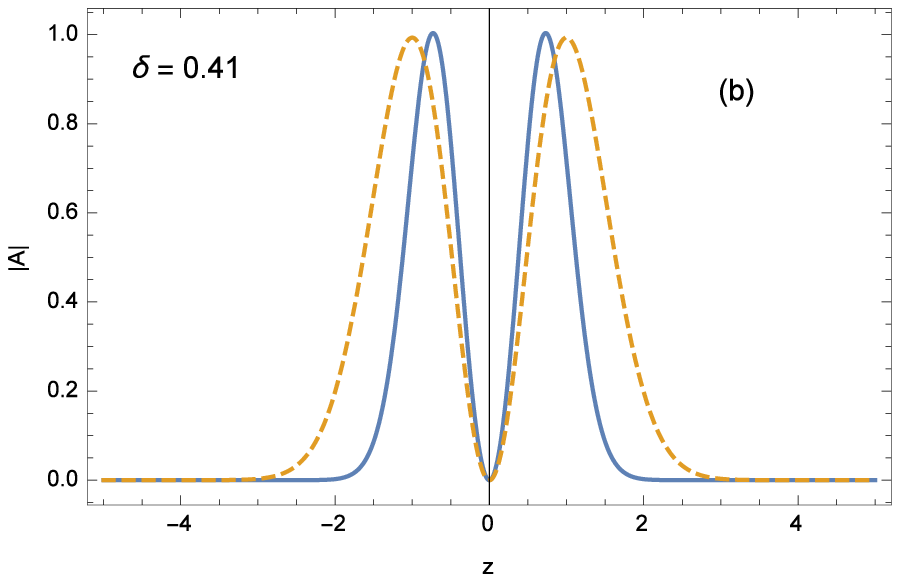}
\end{minipage}

\begin{minipage}[b]{0.5\linewidth}
\includegraphics[width=\textwidth]{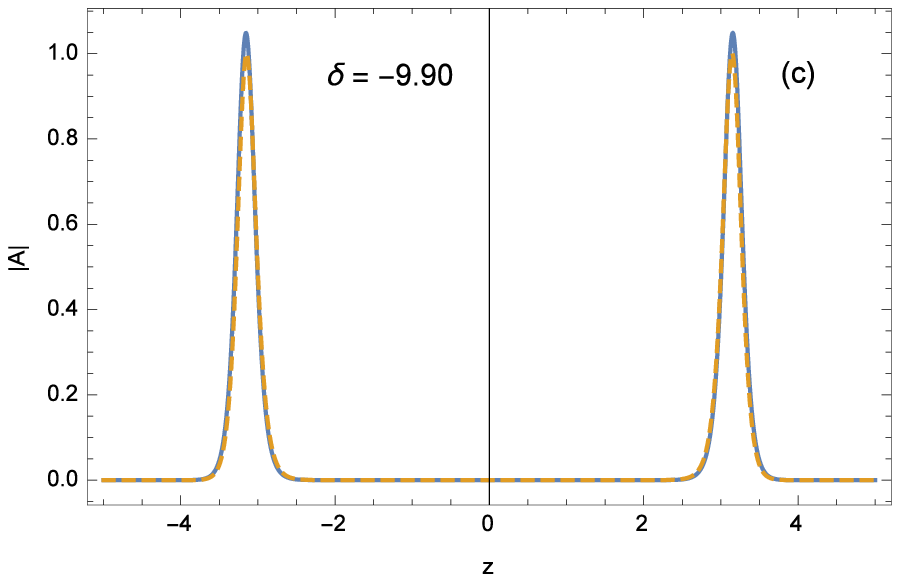}
\end{minipage}%
\begin{minipage}[b]{0.5\linewidth}
\includegraphics[width=\textwidth]{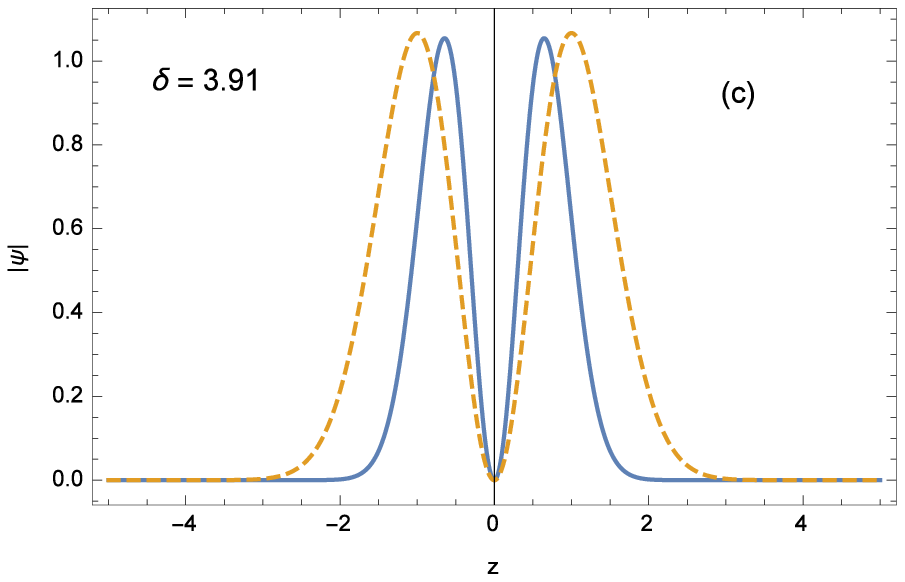}
\end{minipage}
\caption{\label{fig:two} (Color online) Spatial profiles of the exact bisoliton intensity (solid curves in left and right graphs) for $\mu_1=0$ and $\mu_2=0.2$, 
compared with profiles of the even-parity mode (left, dashed lines) and odd-parity mode (right, dashed lines) ansatzes.}
\end{figure*}

\begin{figure*}\centering
\begin{minipage}[b]{0.5\linewidth}
\includegraphics[width=\textwidth]{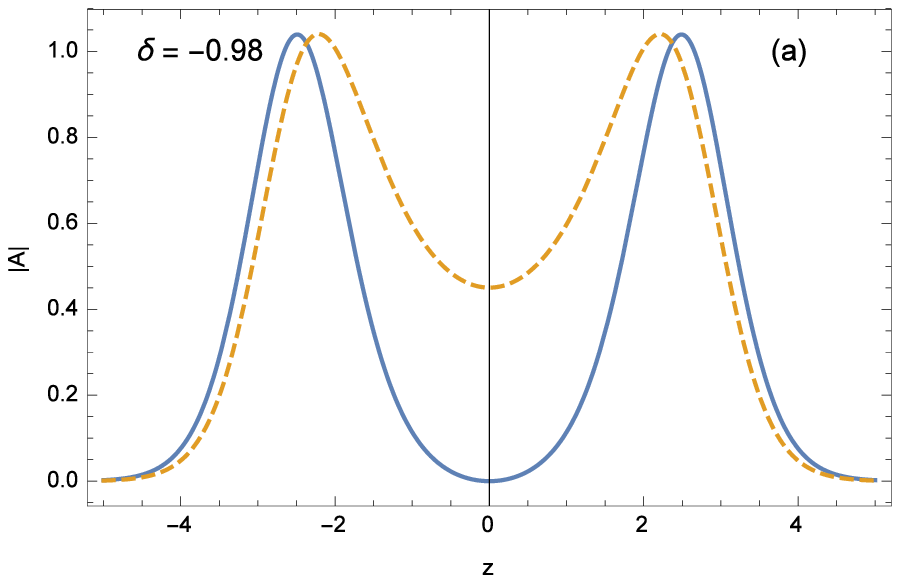}
\end{minipage}%
\begin{minipage}[b]{0.5\linewidth}
\includegraphics[width=\textwidth]{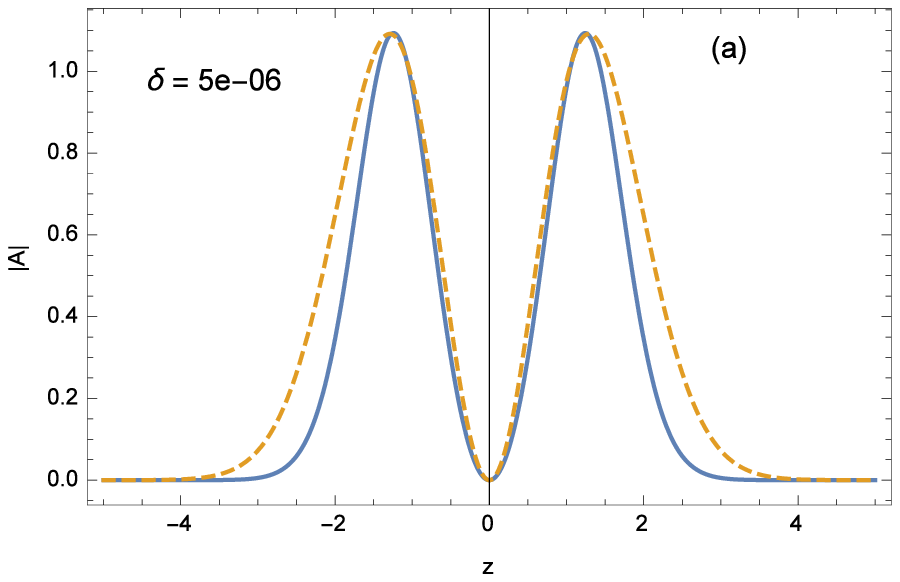}
\end{minipage}

\begin{minipage}[b]{0.5\linewidth}
\includegraphics[width=\textwidth]{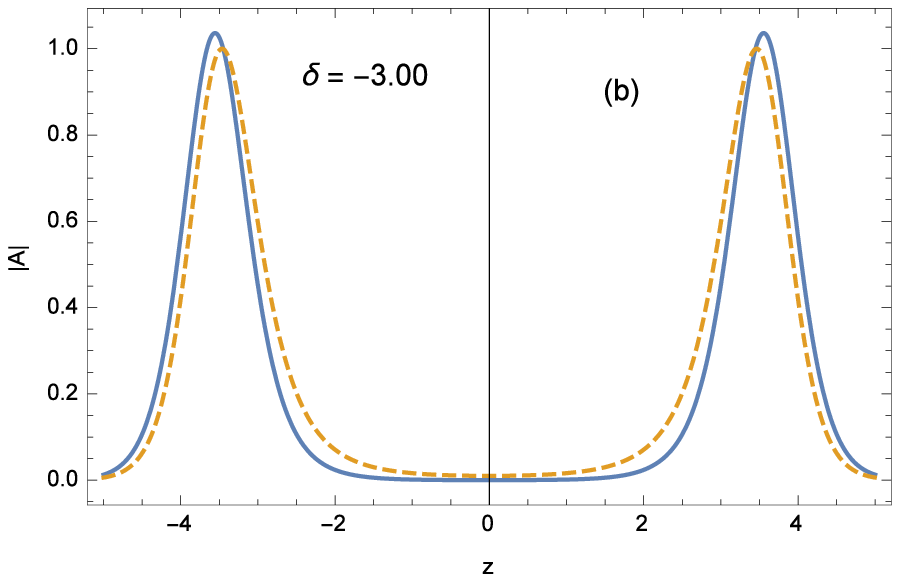}
\end{minipage}%
\begin{minipage}[b]{0.5\linewidth}
\includegraphics[width=\textwidth]{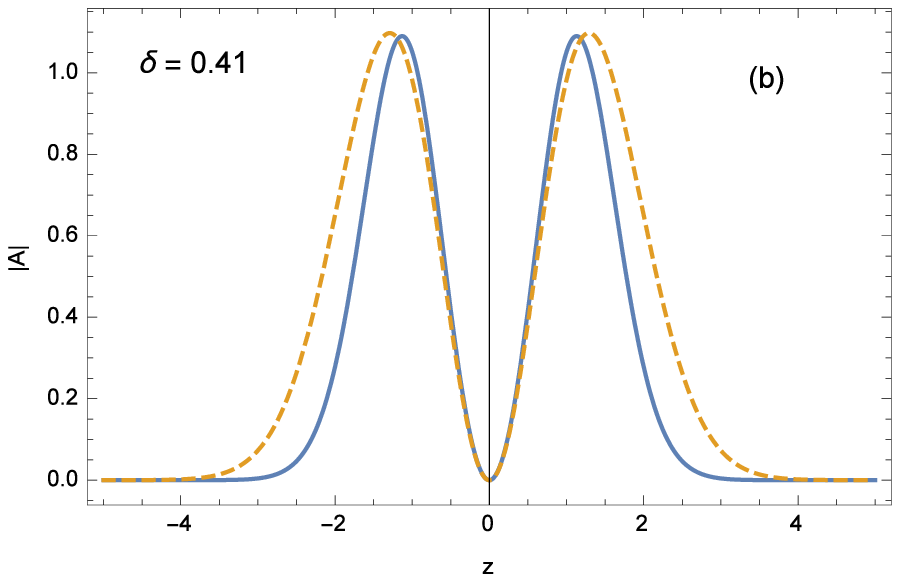}
\end{minipage}

\begin{minipage}[b]{0.5\linewidth}
\includegraphics[width=\textwidth]{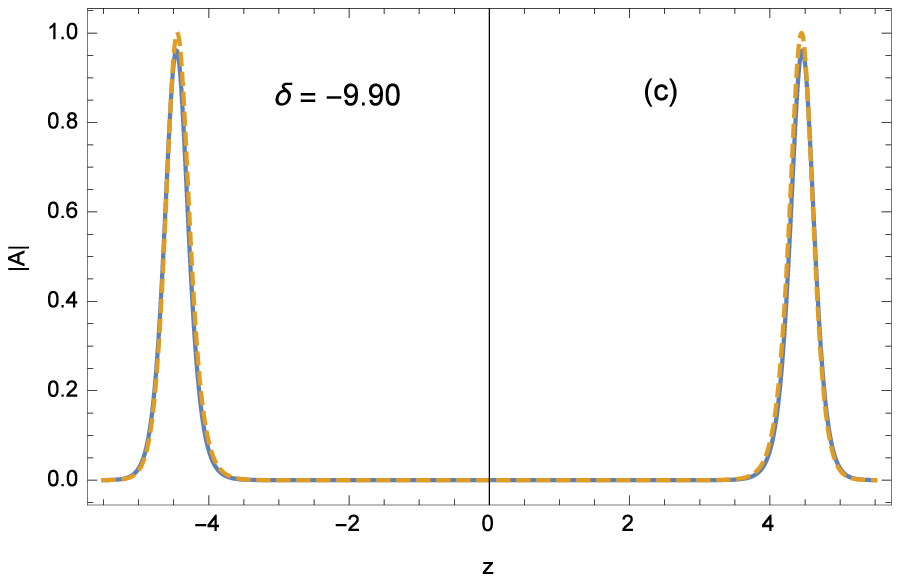}
\end{minipage}%
\begin{minipage}[b]{0.5\linewidth}
\includegraphics[width=\textwidth]{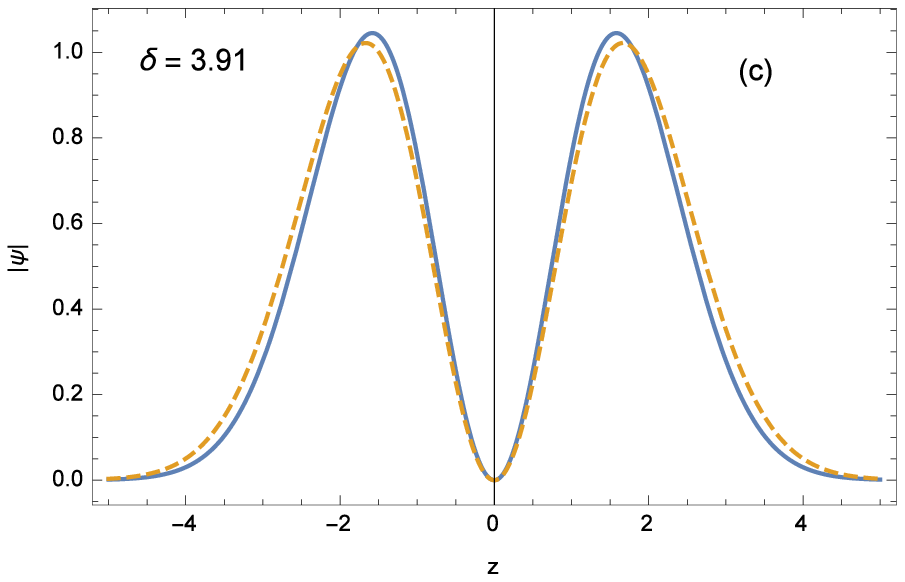}
\end{minipage}
\caption{\label{fig:three} (Color online) Spatial profiles of the exact bisoliton intensity for $\mu_1=0$, $\mu_2=5$, and different values of $\delta$ (solid lines), 
compared with profiles of the equivalent "super-sech"(left, dashed lines) 
and Hermite\text{--}Gaussian (right, dashed lines) ansatzes.}
\end{figure*}

In the previous section, the NLSE~(\ref{eq1}) was solved by following the IST technique, considering a pulse-shaped initial profile describing a bright matter-wave soliton. Regarding this point, the external potential in the NLSE represents a harmonic hump with an asymmetric shape due to gravity. In practice, such a potential can be associated with a harmonic laser barrier that splits a preformed single-pulse soliton into two pulses with an equal tail, width and hence energy. If one of the two post-created pulses is allowed to fall freely under gravity, the gravity-induced acceleration will create an extra potential energy, that varies linearly with the pulse position along the direction of free fall. By measuring the energy difference between the non-accelerated pulse component and the pulse component falling under gravity at some specific position, one can determine exactly the constant of gravity $g$ for atoms of a specific species (i.e. of a well-defined mass $m$) composing a Bose\text{--}Einstein condensate. Thus, the problem considered in this work has a direct application in gravity measurements using matter-wave interferometry, as explained in more detail in refs.~\cite{pezze05,pezze06}.   \\
Although the NIST technique has enabled us find an exact solution to the NLSE~(\ref{eq1}), in recent studies the problem of bisolitonic wave generation and propagation has most often been considered within the framework of variational approaches~\cite{kumar1,Pare99,zang1,konotop1,gardiner1}. The two most common variational ansatzes used in this context are the super-sech and the second-order Hermite\text{--}Gaussian modes. \\
The super-sech mode can be expressed as
  \begin{equation}
 \vert A_{Sup}(z, 0)\vert= A_0(0)\, sech\left[(z - z_c)^2/\sigma_0^2 + \delta_0\right], \label{a31a}
 \end{equation}
while the second-order Hermite\text{--}Gaussian mode is given by
\begin{equation}
\vert A_{HG}(z, 0)\vert= a_1(0)\, z\, e^{-(z - z_c)^2/\sigma_0^2}, 
\label{a31b}
\end{equation} 
where $\sigma_0$, $\delta_0$, $a_1(0)=a_1$, and $A_0(0)=A_0$ are the initial values (i.e., values at $\tau=0$) of the soliton's variational parameters, while $z_c$ is the soliton center-of-mass position. In the spirit of the variational approach, these two predefined solutions are stationary modes, with~(\ref{a31b}) representing the second term of a full Hermite\text{--}Gaussian expansion of the solution to Eq.~(\ref{eq1}) without the linear component in the external potential, treating the nonlinear term as a perturbation~\cite{kaup1}. \\
In fact, the perturbation theory leading to the Hermite\text{--}Gaussian mode~(\ref{a31b}) is the main factor differentiating the two ansatzes. Indeed, the super-sech mode is chosen in such a way as to reproduce the exact NLS single-pulse solution in the absence of external potential. It turns out that the super-sech mode has all the features of a true soliton and it is often referred to as a symmetric (or even-parity) mode~\cite{kumar1,Pare99}. In contrast, the Hermite\text{--}Gaussian ansatz is the solution to a linear eigenvalue problem with a nonlocal nonlinear spatial external potential. For this reason, the Hermite\text{--}Gaussian mode is not a true soliton, despite its localized shape profile. It is now well established that for some values of its variational parameters the Hermite\text{--}Gaussian mode is antisymmetric with respect to its argument $x$, which is why it is sometimes also referred to as an odd-parity mode~\cite{Pare99}. \\
It should be stressed that in our problem, the gravity introduces an extra linear term in the external potential in addition to the quadratic one. Therefore, it is useful to reformulate this variational ansatz. For this purpose we consider a variable change $X=x-X_0$, which transforms the external potential with quadratic plus linear terms given in Eq.~(\ref{eq3b}) into a pure quadratic potential $\tilde{V}(X)$. Doing this, we obtain the same amplitude equation as that solved in ref.~\cite{Pare99} for the temporal Hermite\text{--}Gaussian mode, which in the present context leads to
\begin{equation}
\vert A_{HG}(z, 0)\vert= (a_0 + a_1\, z)\, e^{-(z - z_c)^2/\sigma_0^2}. 
\label{a31c}
\end{equation} 
Concerning the super-sech mode, we shall keep the analytical expression~(\ref{a31a}) since this is actually an arbitrary ansatz as opposed to the Hermite\text{--}Gaussian mode which is the solution to an existing equation. \\
Since our primary goal is to establish a possible consistency between two common variational solutions and the exact soliton solution obtained via the NIST, it will be more practial if we reexpress the exact one-soliton solution given in Eq.~(\ref{eq22}) as 
\begin{equation}
A(z, \tau)= F(z, \tau) \,\exp\lbrace -i\Re{[\varphi(z, \tau)]}\rbrace,  
\label{a32} 
\end{equation}
where $F(z, \tau)=\vert A(z, \tau) \vert$, a real-valued function representing the NIST bisoliton amplitude, follows straight from Eqs.~(\ref{eq23})-(\ref{eq26}) and thus is given by
 \begin{equation}
 F(z, \tau)= \vert A_0(z, \tau)\vert\,  sech\left[(z - 
z_c)^2/\sigma^2(\tau) + \delta(\tau)\right], \label{a33a}
 \end{equation}
with   
 \begin{equation}
 z_c= \mu_1/\mu_2, \label{a33b} 
 \end{equation}
 the bisoliton center-of-mass position
 \begin{equation}
 \sigma^2(\tau)= \frac{g}{\mu_2\,f_{im}(\tau)}, \label{a33c}
\end{equation} 
its spatial width and the quantity
\begin{equation} 
\delta(\tau)=\ln\frac{\vert\vartheta_0\vert}{2f_{im}(\tau)}+4\mu_1\sqrt{\frac{2q}{g}}
\int_0^{\tau}{\Im{[f^2(\tau')]}d\tau'} \label{a33d}
\end{equation} 
accounting for a residual spatial shift in the bisoliton center. Similarly, the quantity $A_0(z, \tau)$ in~(\ref{a33a}), which follows 
from Eq.~(\ref{eq23}) as a prefactor to the sech function, is
 \begin{equation}
 A_0(z, \tau)= -a_1(\tau)z + a_0(\tau), \label{a34}
 \end{equation}
 where
 \begin{eqnarray}
 a_1(\tau)&=& \frac{\vartheta_0\,\mu_2}{R}\sqrt{\frac{q}{2g}}f_{im}(\tau), 
\label{a35a} \\
 a_0(\tau)&=& \frac{\vartheta_0\,\mu_1}{R}\sqrt{\frac{q}{2g}}f_{im}(\tau). 
\label{a35b}
 \end{eqnarray}
In the static regime, we set $\tau=0$ in the above functions, a 
consideration resulring in two essential remarks: 
\begin{enumerate}
\item When we set $\tau=0$ in Eq.~(\ref{a33a}), all functions of $\tau$ in this 
expression become simple parameters. Therefore, we can readily set $\vert 
a_1(0)\vert =a_1$, $\vert a_0(0)\vert =a_0$, $\sigma(0)=\sigma_0$, and 
$\delta(0)=\delta$, such that the parameters $a_1$, $a_0$, $\sigma_0$, and $\delta$ in the 
two variational ansatzes are equal to the equivalent parameters in the exact 
one-soliton solution in the static regime. 
\item According to Eq.~(\ref{eq20}), the quantity 
$f_{im}(0)$ is always positive for finite values of $\eta$. It follows that the
parameter $\delta$, obtained from Eq.~(\ref{a33d}) with $\tau=0$, can take positive and 
negative values depending on the value of $\vartheta_0$ given in formula~(\ref{eq26}). Quite remarkably, while this sign change preserves the bisoliton profile of the exact soliton solution, numerical analysis of the super-sech mode reveals that the change in $\delta$ from positive to negative values gives rise to either a single pulse or a bisoliton. Therefore, we can conclude that in general the bisoliton profile of the propagating super-sech mode is determined by appropriate values of its variational parameter $\delta(t)$, obtained by solving the associated variational equation. 
\end{enumerate}
We start our analysis by comparing intensity profiles of the super-sech mode with those of the exact bisoliton. The three left graphs of figure~\ref{fig:one} represent the amplitudes $\vert 
A(z, \tau=0)\vert$ of the exact bisoliton solution, while the three right graphs are 
the amplitudes of the super-sech ansatz for $\mu_1=0$ 
and $\mu_2=0.2$ and for three different values of $\delta$, i.e., $\delta=0$, 
$3.91$, and $-9.90$. As one can see, while the amplitude of the NIST solution always has a
bisolitonic shape, for positive $\delta$ the super-sech mode has a 
single-pulse shape with an increasingly broad peak as $\delta$ 
increases.
Oppositely, when $\delta$ is decreased in the negative branch, the single-pulse shape
breaks up into a bisoliton, as is apparent in the bottom left 
graph of figure~\ref{fig:one}. The emerging twin pulses are increasingly separated with decreasing 
$\delta$ in the negative branch. Note that in both cases, the peak intensities of the twin pulses are equal. \\
In addition to their equal peak intensities, the two pulses seen in figure~\ref{fig:one} are always symmetric with respect to the fixed center of mass $z_c=0$. This 
feature, together with the equal peak intensities, confers these specific bisolitonic structures with a unique character, which we term the "genuine pulse twinning" state. The main distinctive feature is that the bisoliton energy is always exactly twice the energy of each of the two constituent 
pulses, even in the regime where the twin pulses partially overlap. \\
In our next analysis, we compare the intensity profile of the exact NIST bisoliton with 
the intensity profiles of the Hermite\text{--}Gaussian and super-sech modes. The three left graphs of figures~\ref{fig:two} and~\ref{fig:three} are intensity profiles 
of the NIST soliton (solid curves) and the 
super-sech mode (dashed curves), while the three right graphs are intensity profiles of 
the NIST soliton (solid curves) and the Hermite\text{--}Gaussian mode (dashed 
curves), both plotted in the same graph. The curves in figure~\ref{fig:two} are for $\mu_1=0$ and $\mu_2=0.2$, while the
curves in figure~\ref{fig:three} are for $\mu_1=0$ and a relatively larger value of $\mu_2$ (i.e., $\mu_2=5$). Values of $\delta$ 
are selected in a range where both the super-sech and Hermite\text{--}Gaussian modes display
bisoliton shapes.   \\ 

\begin{figure*}\centering
\begin{minipage}[b]{0.5\linewidth}
\includegraphics[width=\textwidth]{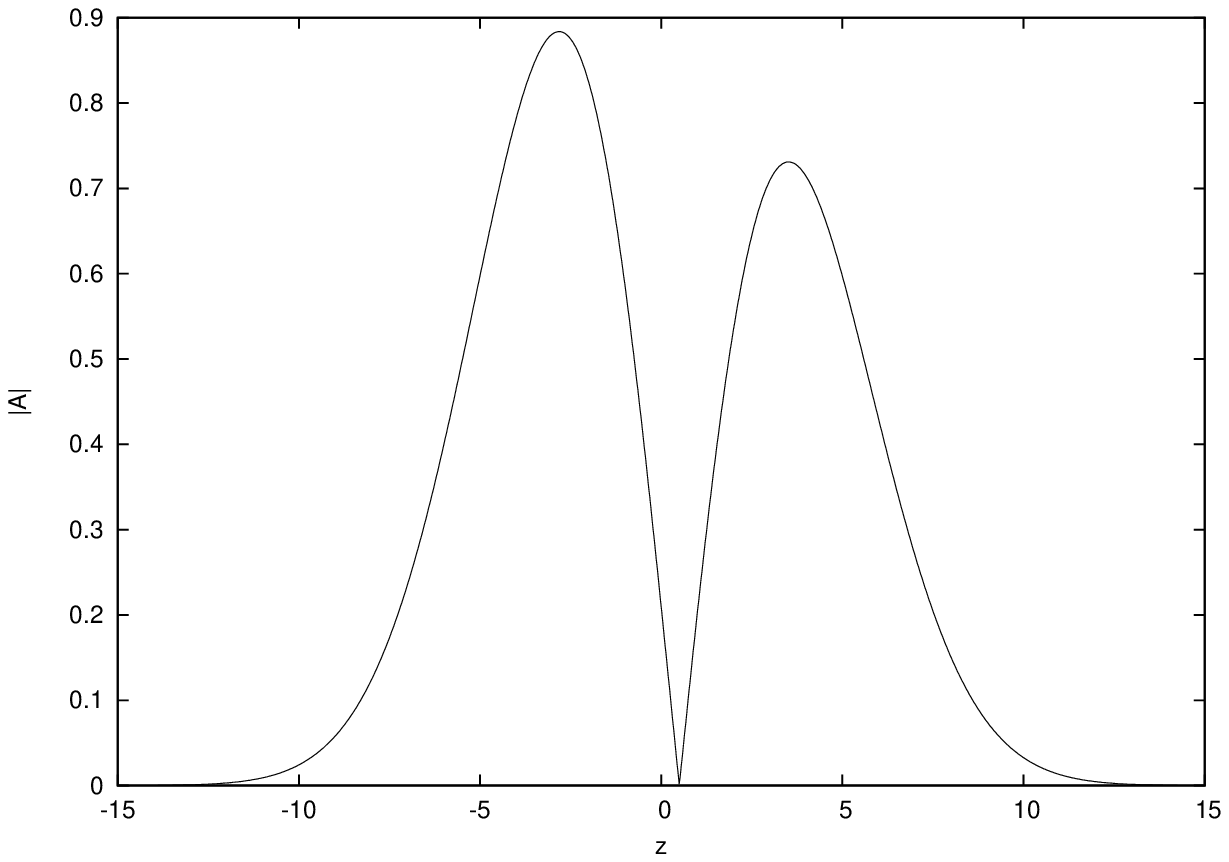}
\end{minipage}%
\begin{minipage}[b]{0.5\linewidth}
\includegraphics[width=\textwidth]{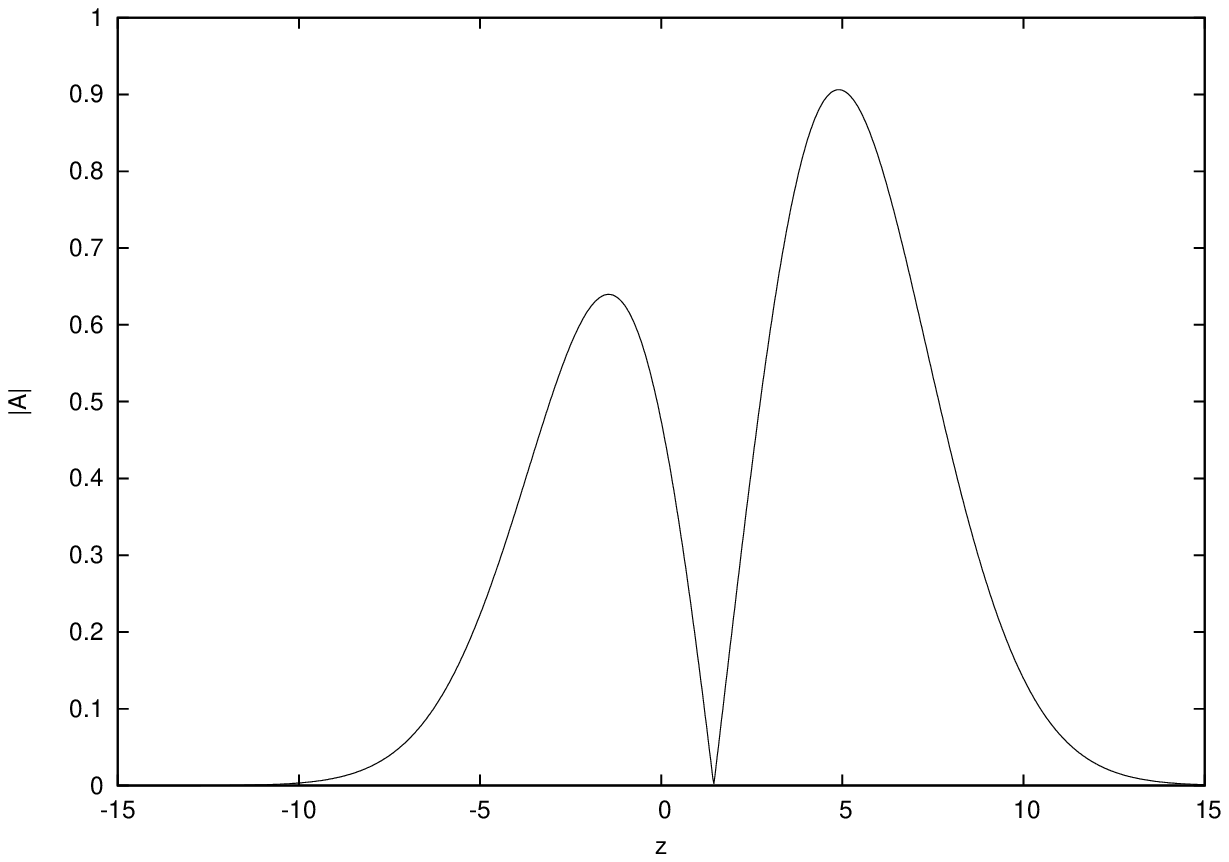}
\end{minipage}\\

\begin{minipage}[b]{0.5\linewidth}
\includegraphics[width=\textwidth]{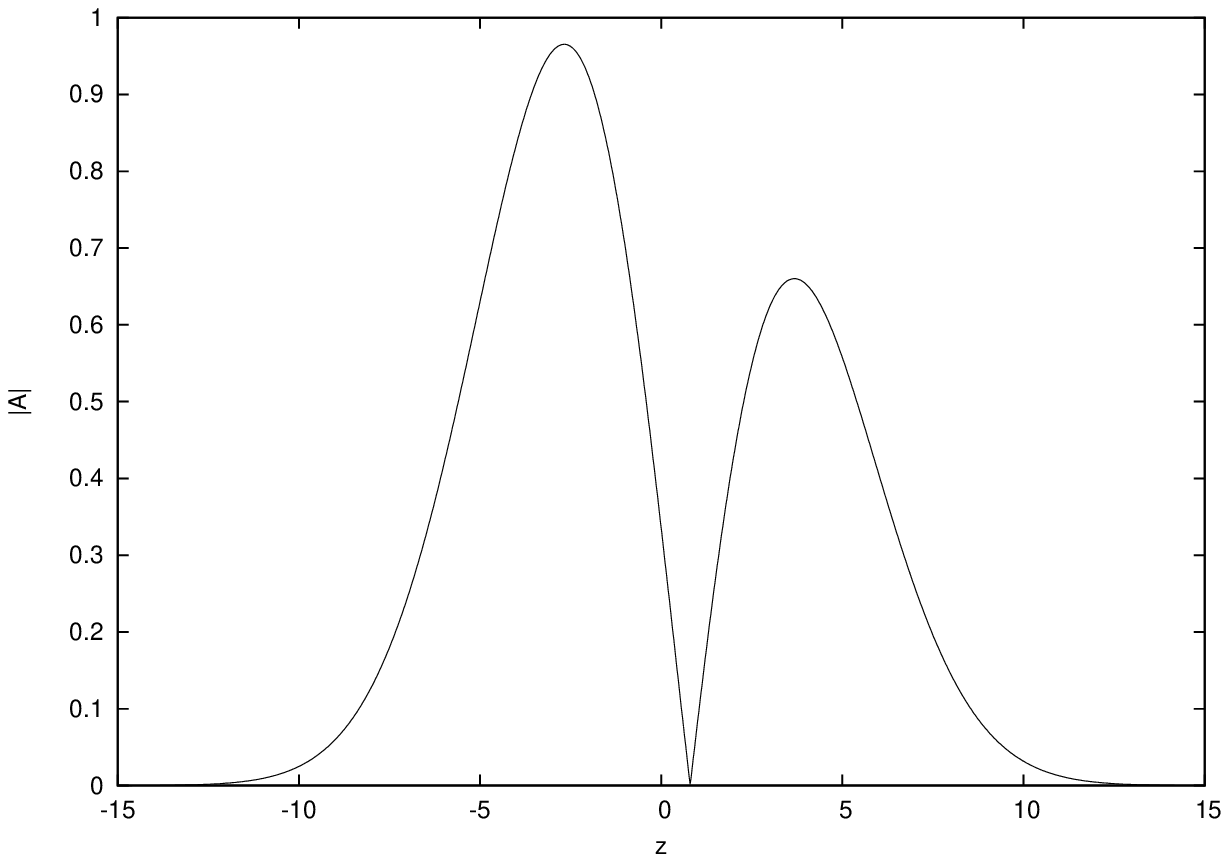}
\end{minipage}%
\begin{minipage}[b]{0.5\linewidth}
\includegraphics[width=\textwidth]{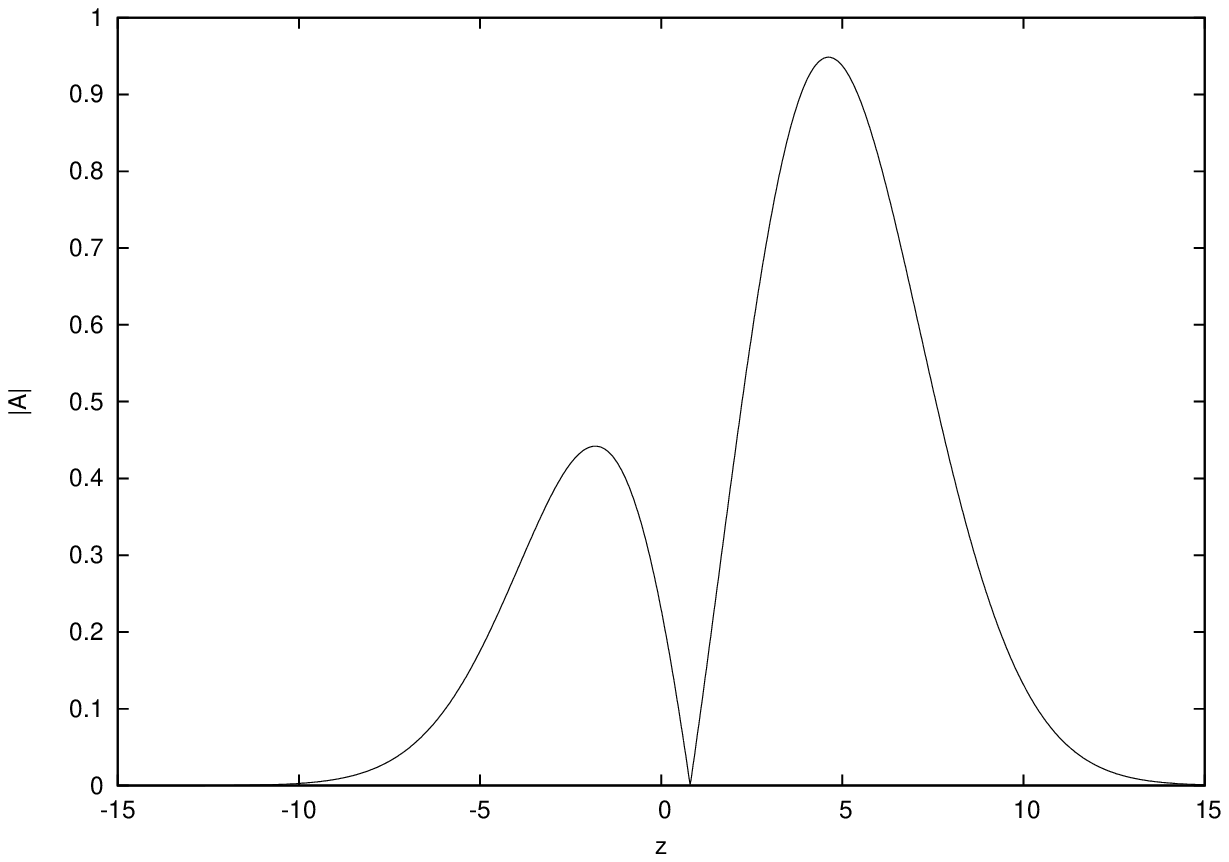}
\end{minipage}
\caption{\label{fig:four} (Color online) Spatial profiles of the Hermite\text{--}Gaussian mode intensity for $\mu_2=0.2$ and different values of $\mu_1$ selected as follows. Left graphs: $\mu_1=-0.1$ (top graph) and $\mu_1=-0.5$ (bottom graph). Right graphs: $\mu_1=0.1$ (top graph) and $\mu_1=0.5$ (bottom graph).}
\end{figure*}

According to the left graphs of figures.~\ref{fig:two} and~\ref{fig:three}, when $\delta$ is decreased in the 
negative branch, the NIST soliton and the super-sech mode tend to two qualitatively identical bisolitonic structures. In contrast to this behavior, the right graphs suggest a tendency to 
identical bisoliton profiles provided $\delta$ is increased in the positive 
branch. Thus, the ranges of $\delta$ values in which the 
even-parity and odd-parity bisolitons are both consistent with the shape profiles of the exact NIST soliton do not match. \\
In the above analysis, we assumed $\mu_1=0$ corresponding to the context of beam splitting without post-gravitational acceleration of the output matter-wave twin pulses. Now we turn to the case that atoms experience gravity. For this case we have shown above that it is possible to mathematically account for the effect of gravity on the analytical expression for the Hermite\text{--}Gaussian mode~(see Eq. \ref{a31c}), but this is not possible for the super-sech ansatz, which is actually an arbitrarily chosen variational ansatz. Therefore, in this second analysis, we focus on the comparision of the Hermite\text{--}Gaussian mode, now given by Eq.~(\ref{a31c}), with the exact one-soliton solution. In figure~\ref{fig:four} we show the intensity profile of the Hermite\text{--}Gaussian mode for $\mu_2=0.2$ with $\mu_1=-0.1$, $\mu_1=-0.8$ (two left graphs), and $\mu_1=0.1$, $\mu_1=0.8$ (two right graphs) in order to see the effects of the linear term in the external potential on the Hermite\text{--}Gaussian quasi-bisoliton shape.  

\begin{figure*}\centering
\begin{minipage}[b]{0.5\linewidth}
\includegraphics[width=\textwidth]{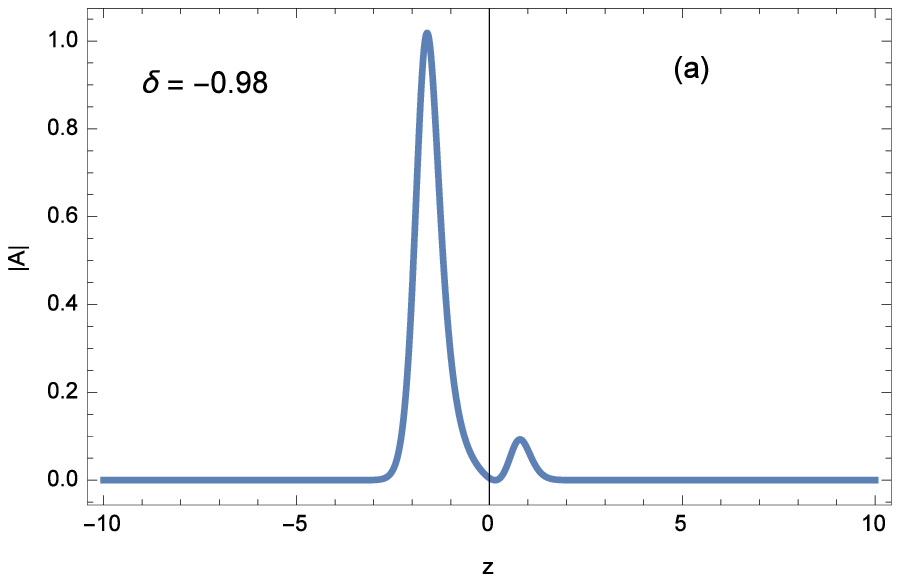}
\end{minipage}%
\begin{minipage}[b]{0.5\linewidth}
\includegraphics[width=\textwidth]{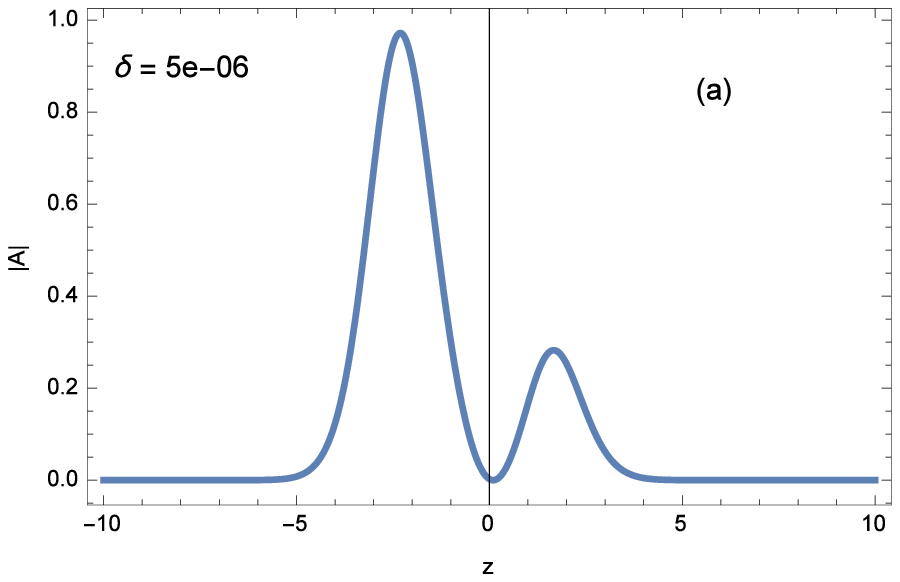}
\end{minipage}

\begin{minipage}[b]{0.5\linewidth}
\includegraphics[width=\textwidth]{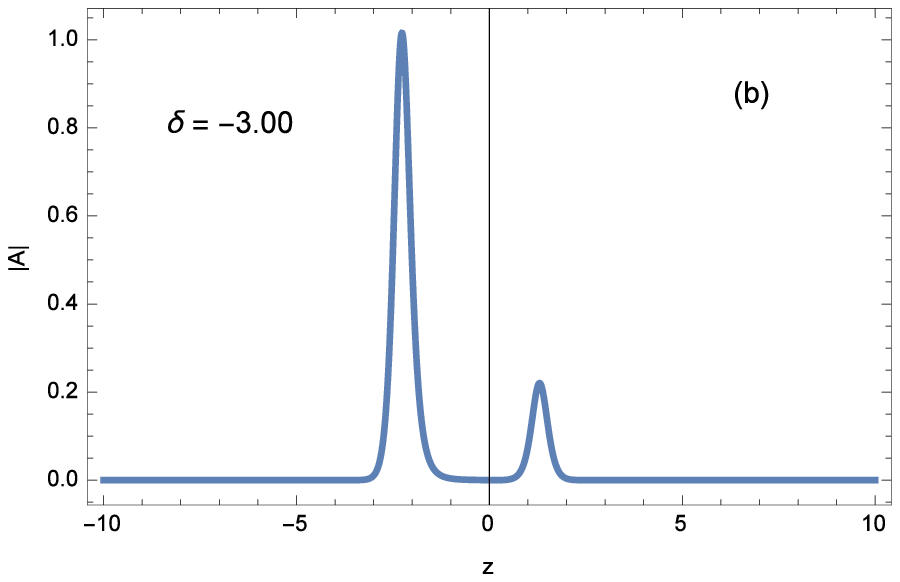}
\end{minipage}%
\begin{minipage}[b]{0.5\linewidth}
\includegraphics[width=\textwidth]{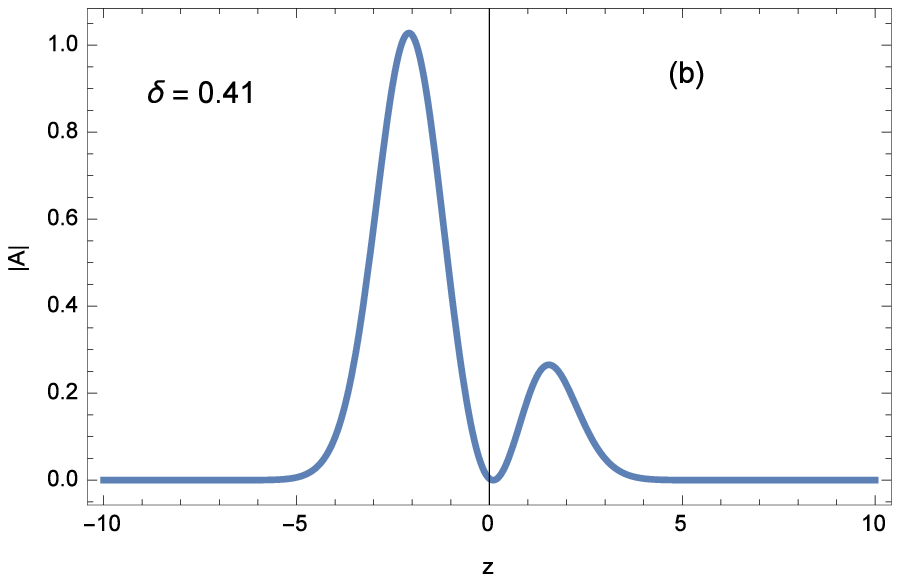}
\end{minipage}

\begin{minipage}[b]{0.5\linewidth}
\includegraphics[width=\textwidth]{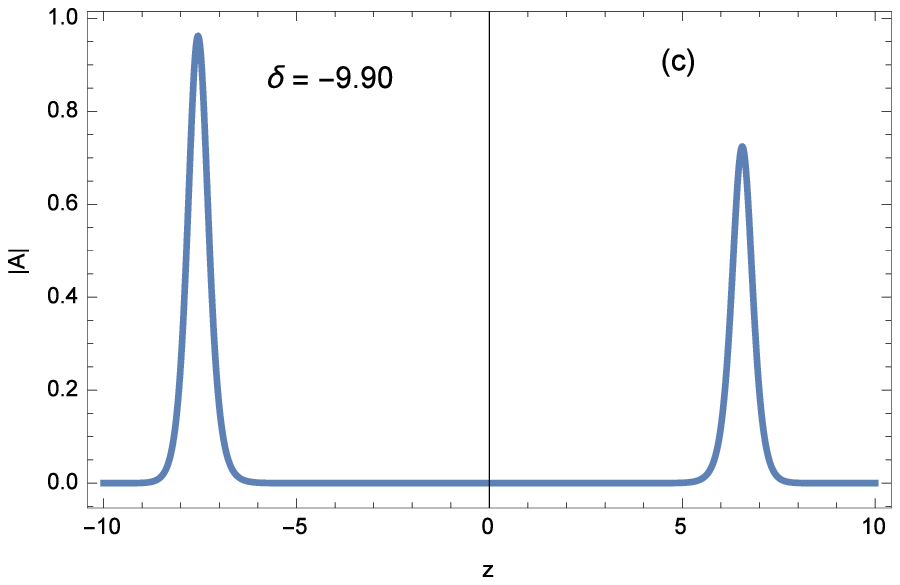}
\end{minipage}%
\begin{minipage}[b]{0.5\linewidth}
\includegraphics[width=\textwidth]{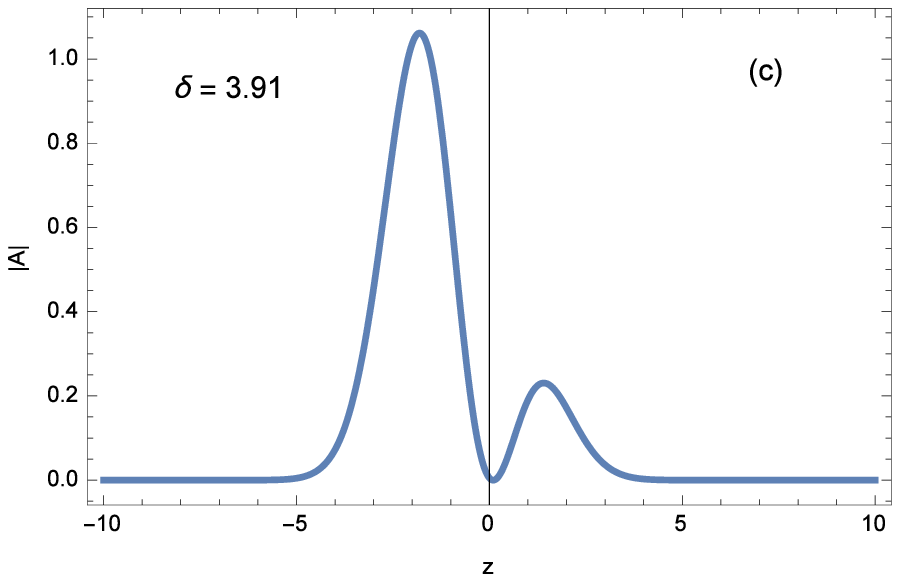}
\end{minipage}
\caption{\label{fig:four1} (Color online) Spatial profiles of the exact bisoliton intensity for 
$\mu_1=0.1$, $\mu_2=0.2$, and three distinct negative (left) and 
positive (right) values of $\delta$.}
\end{figure*}

\begin{figure*}\centering
\begin{minipage}[b]{0.5\linewidth}
\includegraphics[width=\textwidth]{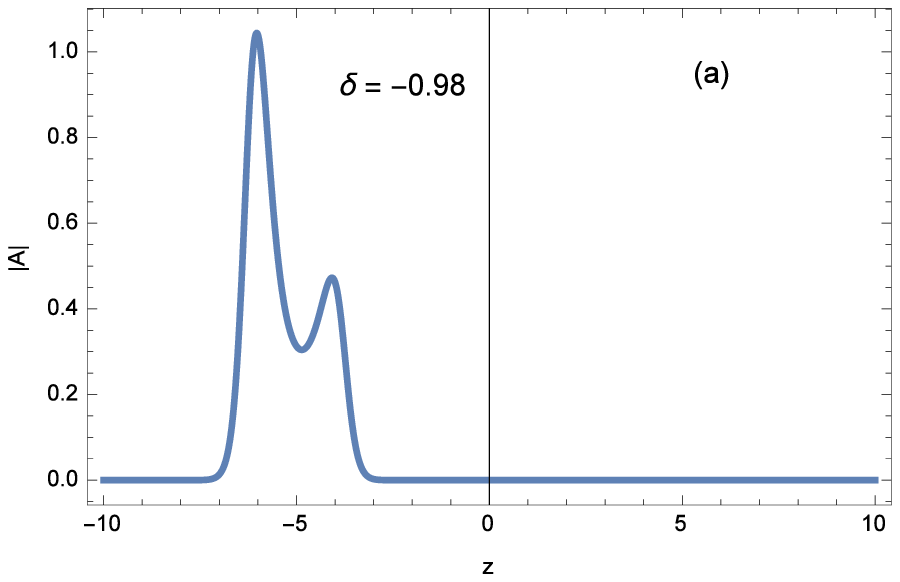}
\end{minipage}%
\begin{minipage}[b]{0.5\linewidth}
\includegraphics[width=\textwidth]{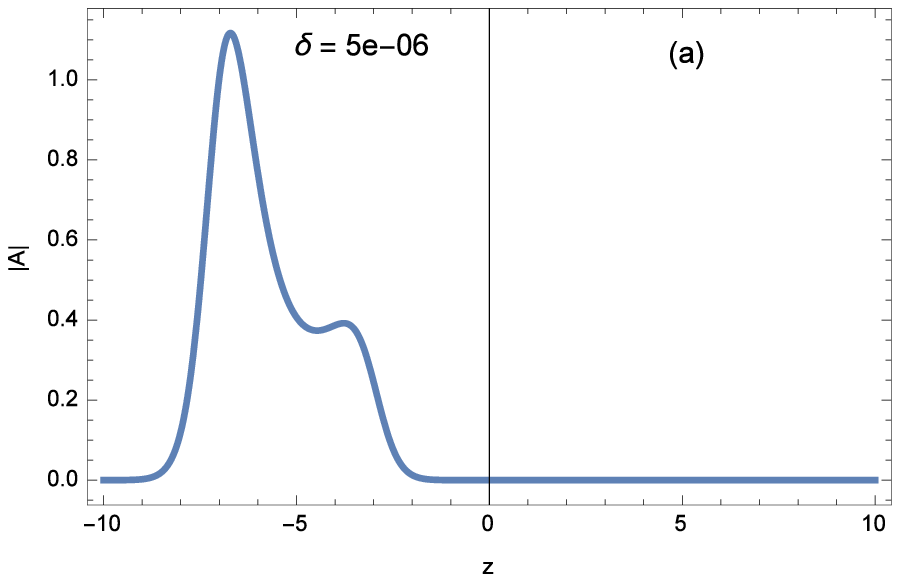}
\end{minipage}

\begin{minipage}[b]{0.5\linewidth}
\includegraphics[width=\textwidth]{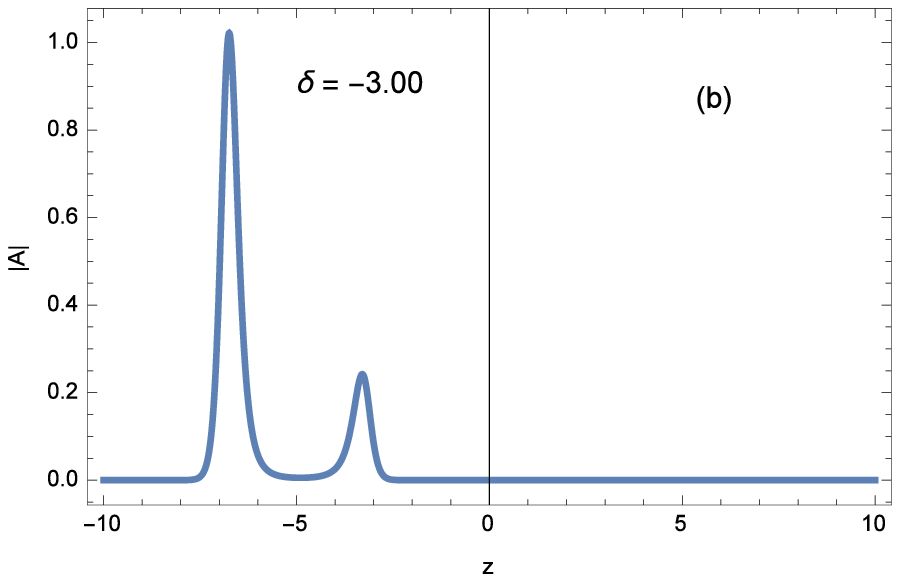}
\end{minipage}%
\begin{minipage}[b]{0.5\linewidth}
\includegraphics[width=\textwidth]{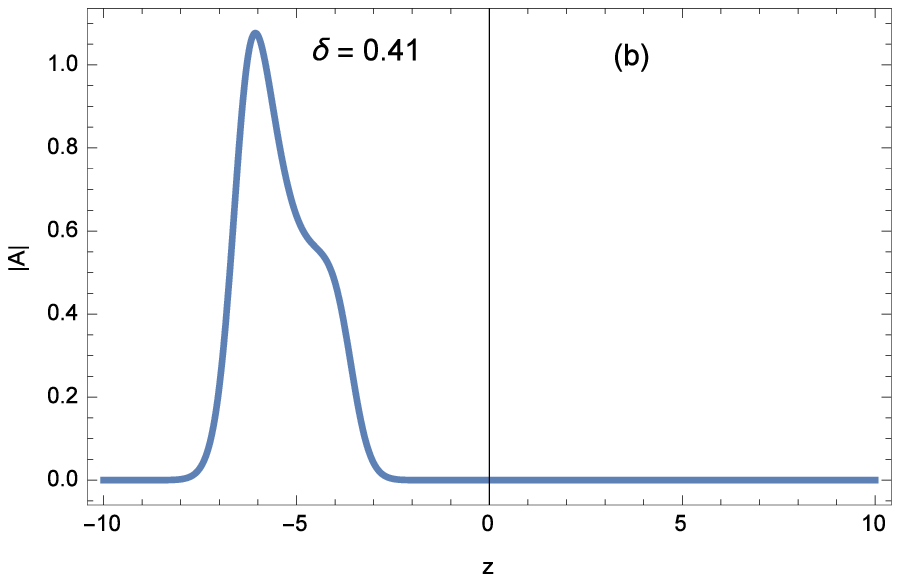}
\end{minipage}

\begin{minipage}[b]{0.5\linewidth}
\includegraphics[width=\textwidth]{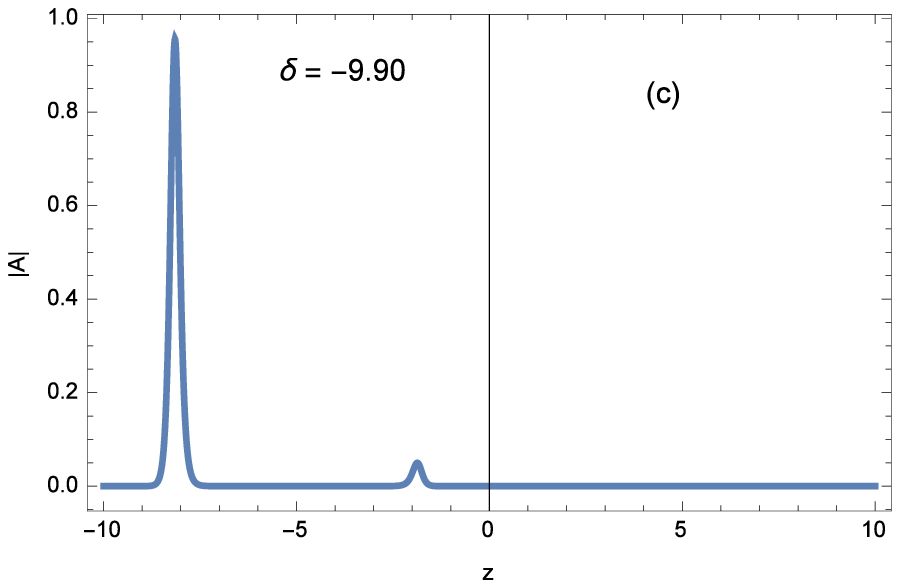}
\end{minipage}%
\begin{minipage}[b]{0.5\linewidth}
\includegraphics[width=\textwidth]{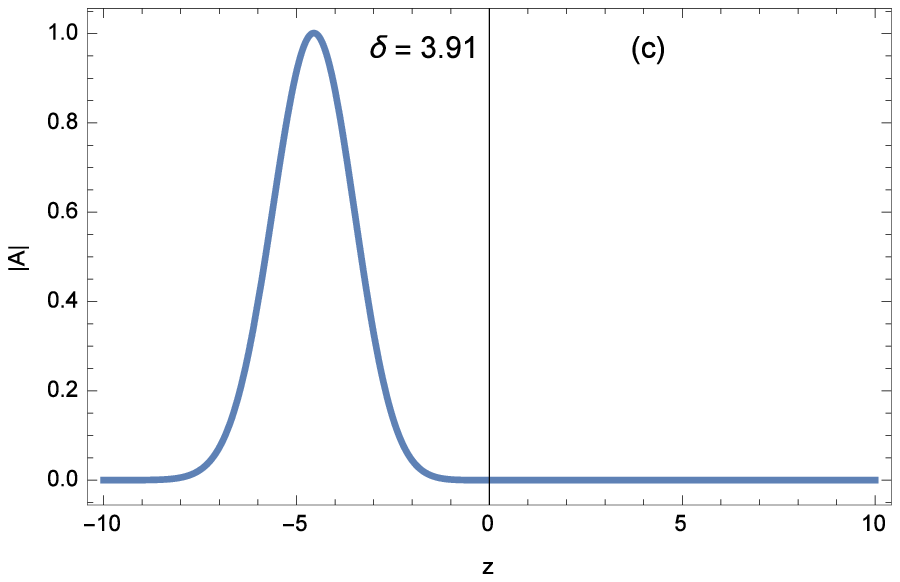}
\end{minipage}
\caption{\label{fig:five} (Color online) Spatial profiles of the exact bisoliton intensity for $\mu_1=1$, $\mu_2=0.2$, and three distinct negative (left) and positive (right) values of $\delta$.}
\end{figure*}

For the exact bisoliton, in figures~\ref{fig:four1} and~\ref{fig:five} we plotted its intensity profiles for $\mu_1=0.1$, 
$\mu_2=0.2$, and $\mu_1=1$, $\mu_2=0.2$, respectively, for three distinct negative (left graphs) and positive (right graphs) values of $\delta$. 

When both $\mu_1$ and $\mu_2$ are relatively small (Fig.~\ref{fig:four1}), the bisoliton feature survives but the two 
pulses have different widths and tails. In addition, the center positions 
of the twinned pulses are no longer symmetric with respect to $z = 0$, as was the case in the absence of a gravity potential. 
Instead we have two asymmetric pulses. On the other hand, when the linear coefficient 
$\mu_1$ is relatively large compared with the quadratic coefficient $\mu_2$ (Fig.~\ref{fig:five}), the bisoliton 
shape is destroyed for certain values of $\delta$. Given that for the exact bisoliton, $\delta$ depends mainly on 
the characteristic parameters of the matter-wave evolution equation, we can exqctly determine the existence condition for a matter-wave bisoliton structure. These features of the exact bisoliton solution, namely, the shift of pulse center positions with varying characteristic parameters of the system, cannot be accounted for by the Hermite\text{--}Gaussian mode as is apparent in figure~\ref{fig:four}. Indeed, figure~\ref{fig:four} clearly shows that although the intensities of the two pulses are not equal, their amplitudes persistently overlap for all values of $\delta$.

\section{Numerical Simulations}
\begin{figure*}\centering
\begin{minipage}[b]{0.52\textwidth}
\includegraphics[width=2.7in,height=2.8in]{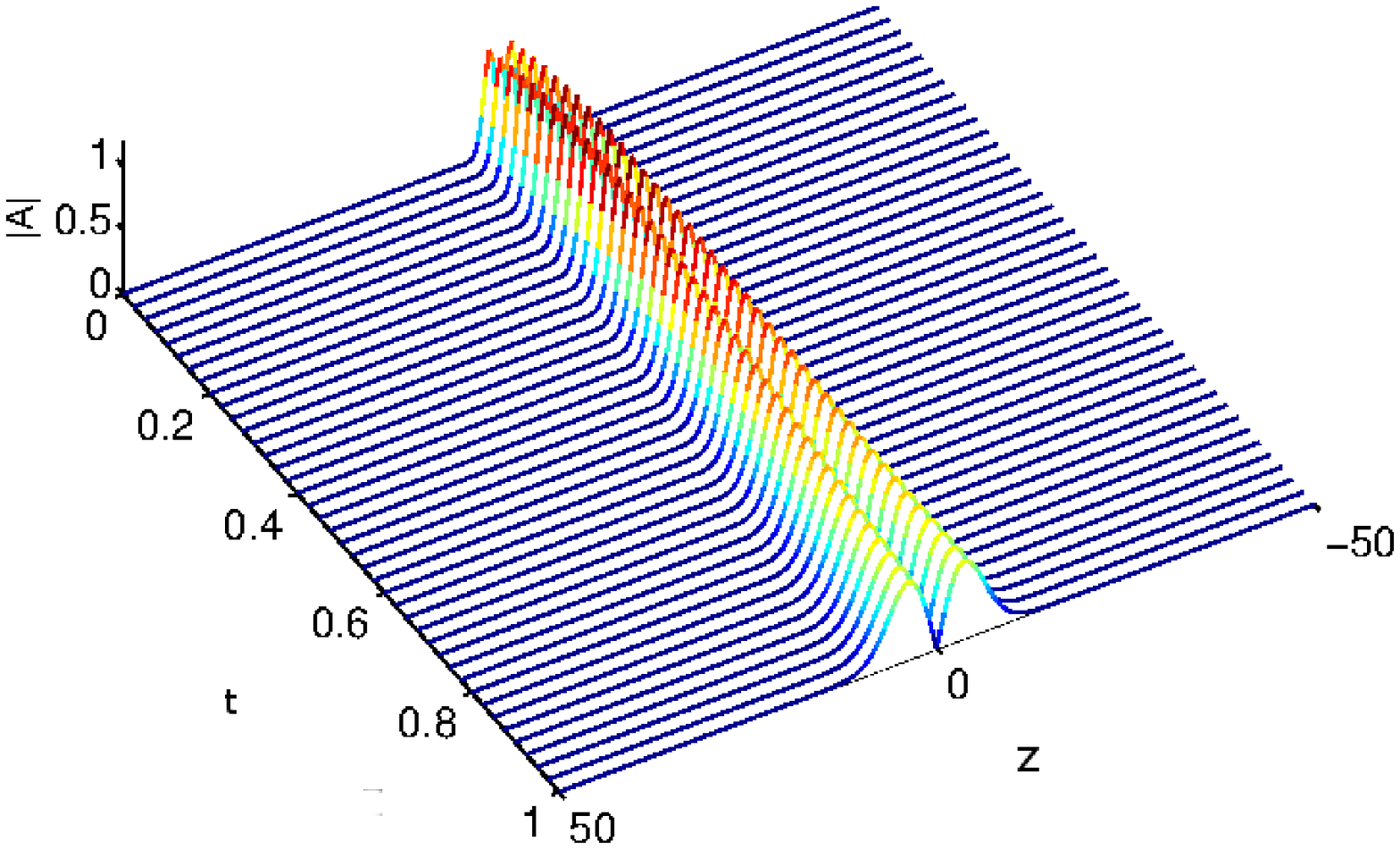}
\end{minipage}%
\begin{minipage}[b]{0.52\textwidth}
\includegraphics[width=2.7in,height=2.8in]{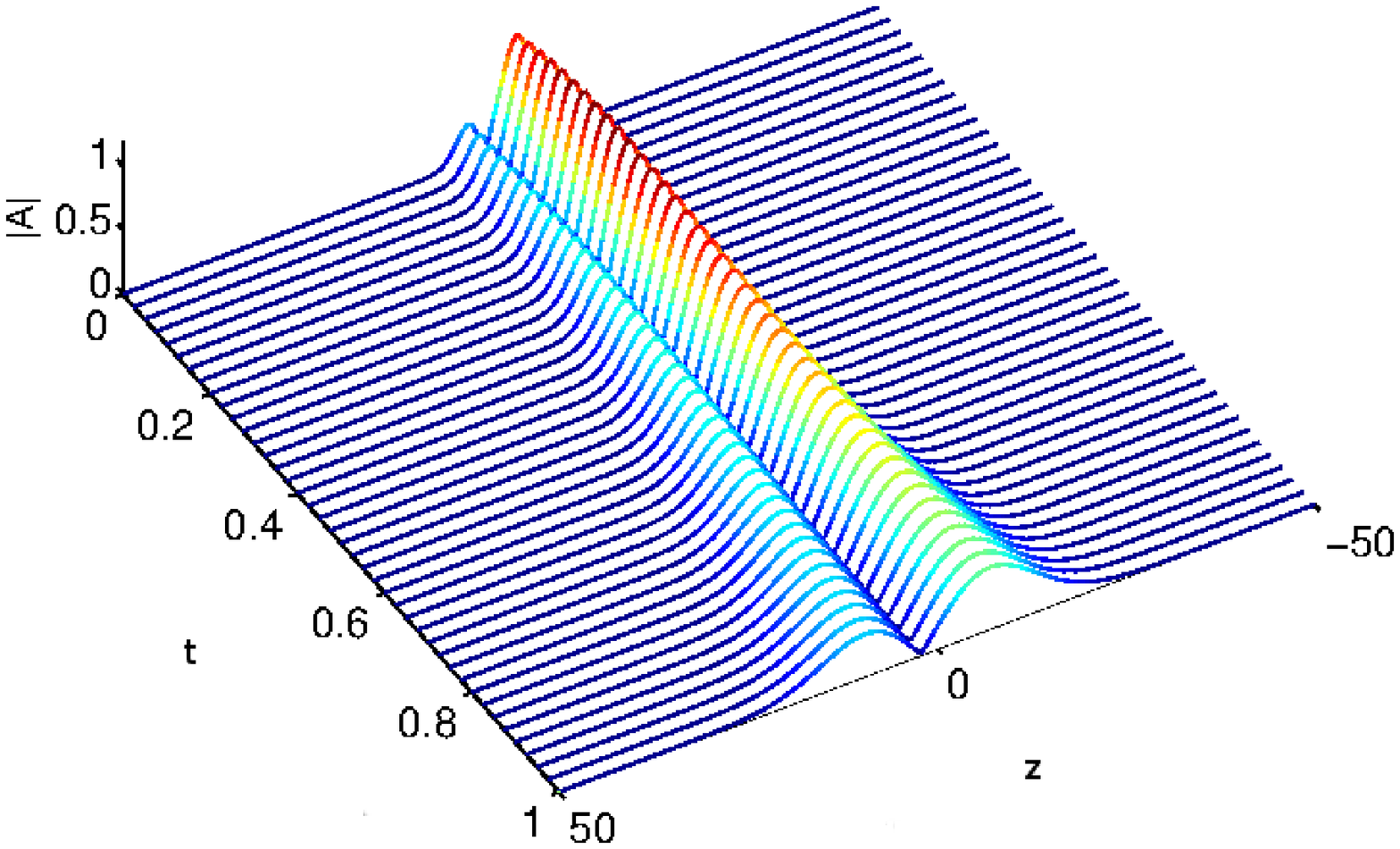}
\end{minipage}
 \caption{\label{fig:six} (Color online) Numerical profiles of the matter-wave field intensity for $\delta=0$. Left graph: $\mu_1=0$, $\mu_2=0.2$. Right graph: $\mu_1=0.1$, $\mu_2=0.2$.}
\end{figure*}

\begin{figure*}\centering
\begin{minipage}[b]{0.52\textwidth}
\includegraphics[width=2.7in,height=2.8in]{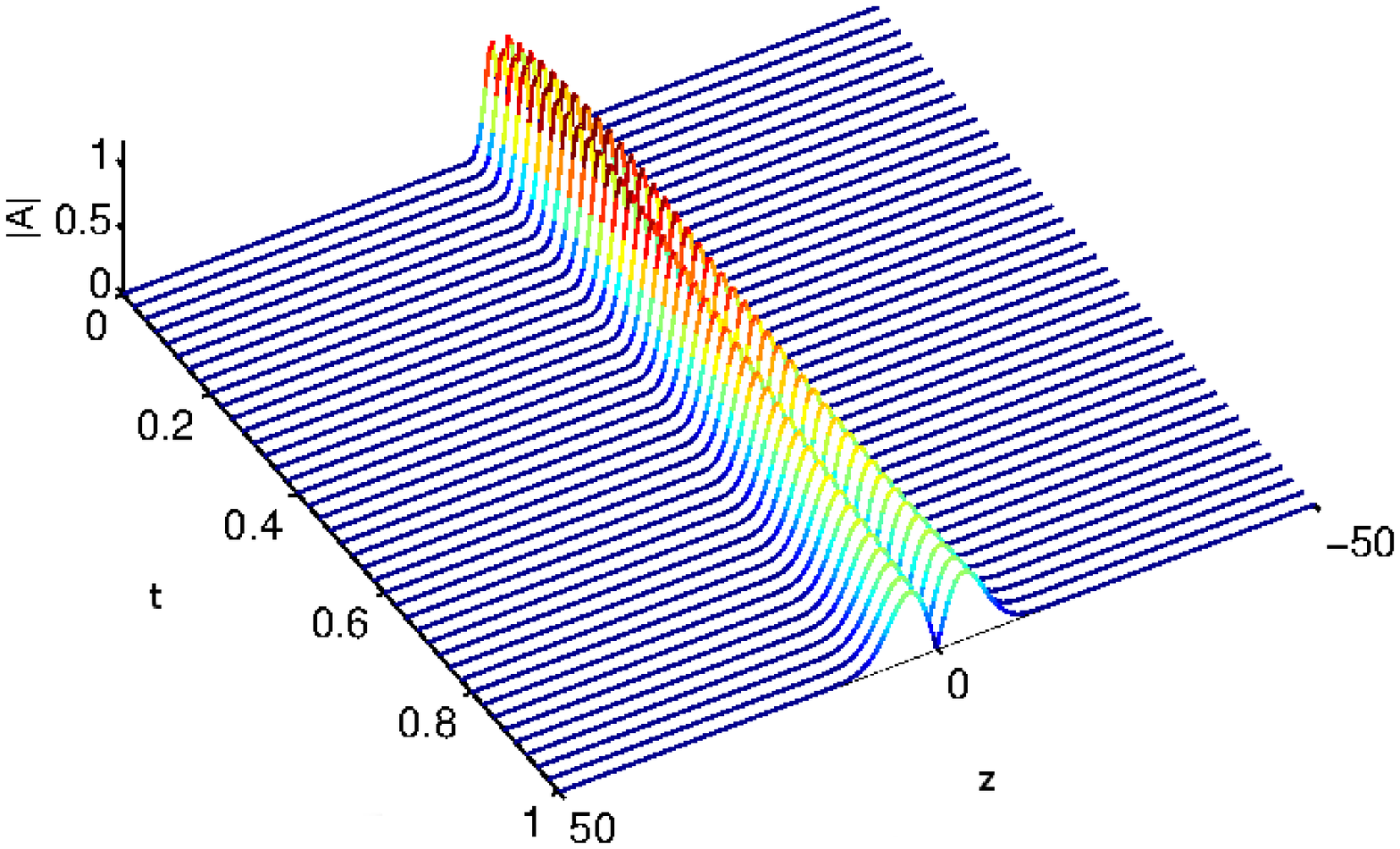}
\end{minipage}%
\begin{minipage}[b]{0.52\textwidth}
\includegraphics[width=2.7in,height=2.8in]{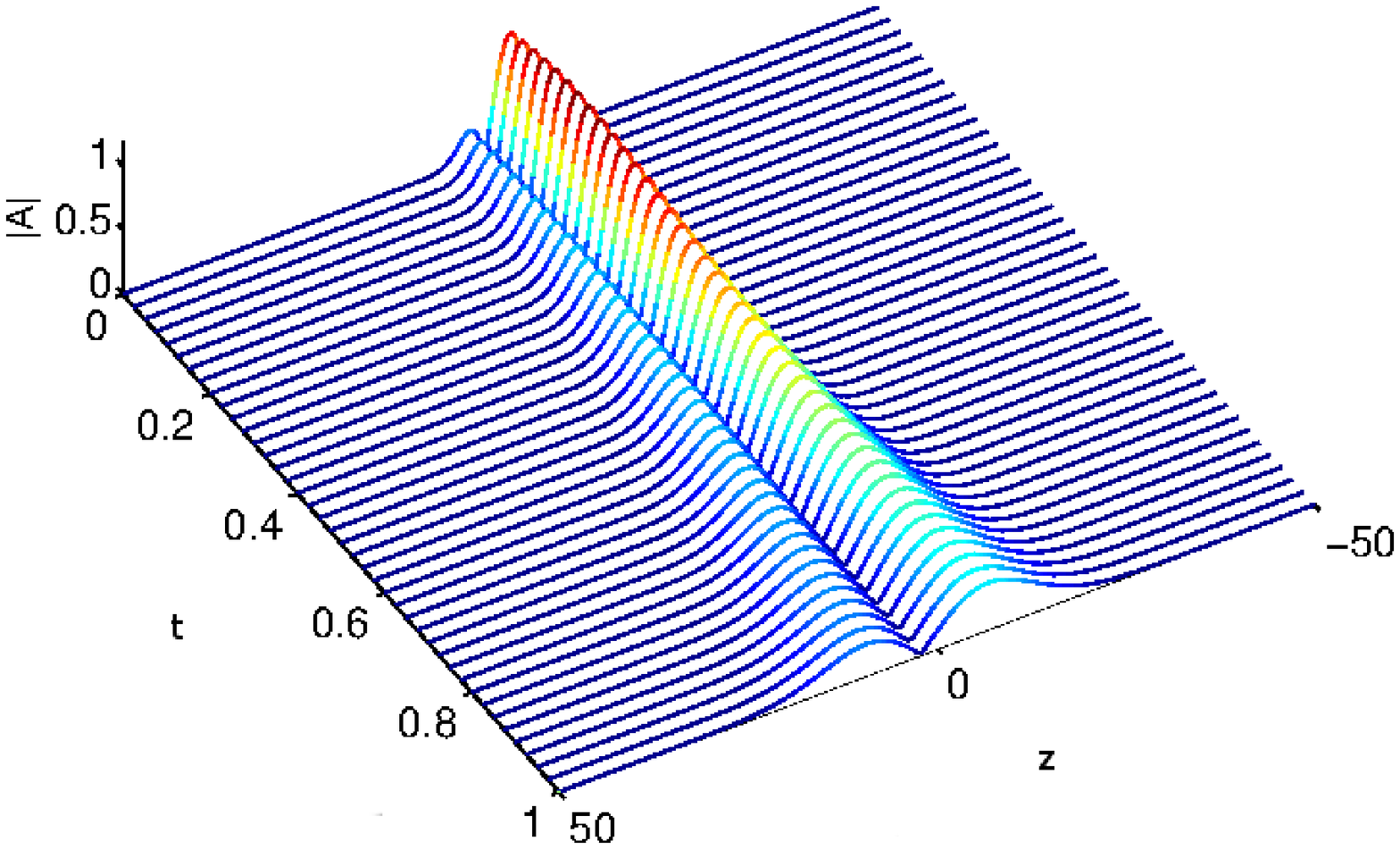}
\end{minipage}
\caption{\label{fig:seven}  Numerical profiles of the matter-wave field intensity for $\delta=2.5$. Left graph: $\mu_1=0$, $\mu_2=0.2$.
Right graph: $\mu_1=0.1$, $\mu_2=0.2$.}
\end{figure*}

\begin{figure*}\centering
\begin{minipage}[b]{0.52\textwidth}
\includegraphics[width=2.7in,height=2.8in]{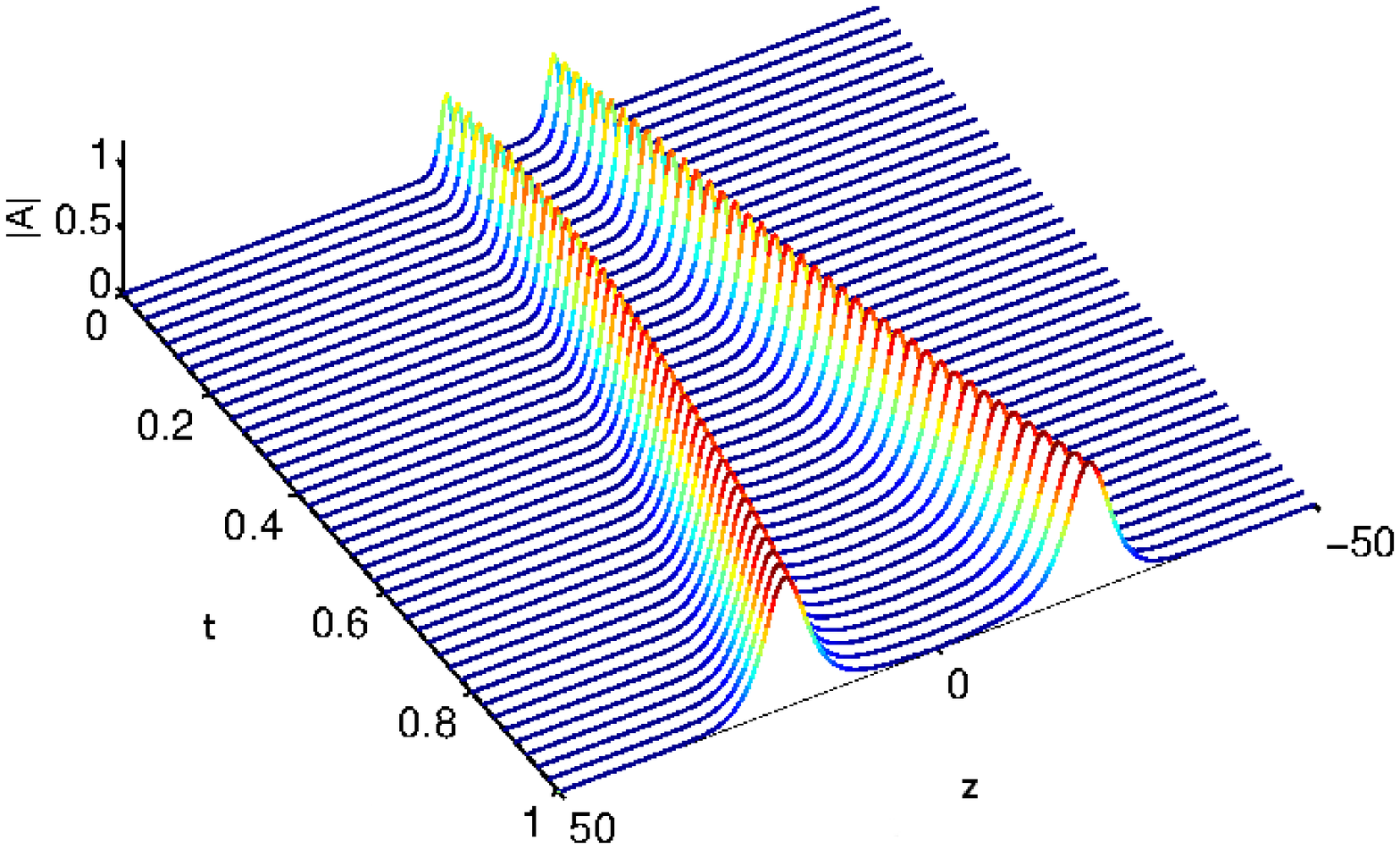}
\end{minipage}%
\begin{minipage}[b]{0.52\textwidth}
\includegraphics[width=2.7in,height=2.8in]{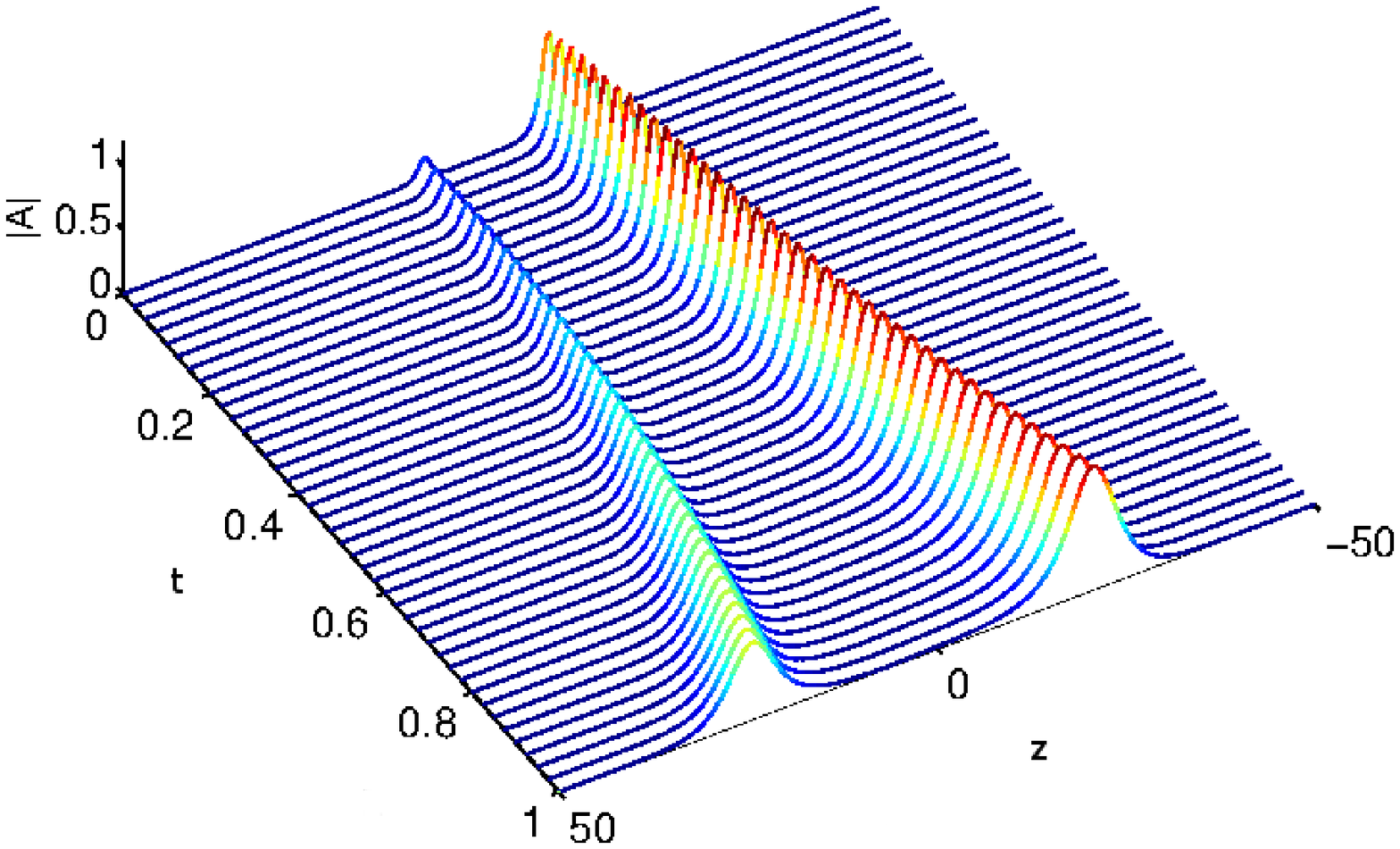}
\end{minipage}
\caption{\label{fig:eight} (Color online) Numerical profiles of the matter-wave field intensity for $\delta=-2.5$. Left graph: $\mu_1=0$, $\mu_2=0.2$. Right graph: $\mu_1=0.1$, $\mu_2=0.2$.}
\end{figure*}

In the previous section, an extensive analysis of the consistency between the exact soliton solution to the GP Eq. (\ref{eq4}), on one hand, and two variational ansatzes to the same equation, on the other hand, was carried out. We found that the two variational fields agree qualitatively with the exact solution only in restricted ranges of their characteristic parameters. Most importantly, the parameter values for which the Hermite\text{--}Gaussian and the super-sech modes display the required double-pulse profile are distinct. However, the exact solution is always a double-pulse structure, with the pulse center positions and intensities changing with the linear and quadratic coefficients of the external potential.\\
Nevertheless, the analysis in the previous section considered only static profiles of the fields. This is because the dynamical properties of the two variational fields can be determined only by solving appropriate variational equations, which is beyond the scope of the present work. In fact, our main interest was the consistency of the shape profiles of the three solutions. Concerning the system dynamics, the most relevant information about the behaviors of the matter-wave field is expected from the exact solution to the equation of motion, i.e., the GP equation. In this respect, we simulated the GP equation (\ref{eq4}) by following a split-step scheme in the time domain, assuming that the spatial profile of the field is given by Eq. (\ref{a32}) at $\tau=0$. With this consideration, we can vary the characteristic parameter $\delta(0)=\delta$, in addition to the linear and quadratic coefficients of the external position. \\
Figures~\ref{fig:six} ($\delta=0$),~\ref{fig:seven} ($\delta=2.5$), and~\ref{fig:eight} ($\delta=-2.5$) display profiles of the field intensity in space-time coordinates for some sets of values of characteristic model parameters.
Figures~\ref{fig:six}, which corresponds to the numerical results for $\delta=0$, shows that in the absence of gravity, the matter-wave field has a symmetric double-pulse structure throughout its propagation time. When $\mu_1=0$ (i.e., in the absence of gravity), the two co-propagating pulses form a twin pair with the same intensity, but the separation between their peaks seems to change only weakly. The same behavior emerges in figure~\ref{fig:seven} for a nonzero but positive value of $\delta$. However, when $\delta$ is negative, the peak separation manifestly varies in time. More explicitely, the two pulses move away from each other with time, while the heights of their peaks increase. Note that the quantity $\delta$ for the exact soliton solution to the GP equation is defined in Eq. (\ref{a33d}) it is indeed negative by virtue of the first term in the right-hand side of Eq. (\ref{a33d}) if $\vartheta_0$ is very small compared with $f_{im}(\tau=0)$. 

\section{Discussion and Concluding Remarks}
Matter-wave interferometry is one of the most active current fields of investigation in quantum metrology, because of its great advantage related to the finite mass of the field constituents that permits the probing of quantum properties of matter, and force fields such as gravity, matter-matter interactions and so forth. \\
In a double-slit experiment involving a matter-wave field to measure the gravity force, for instance, one can assume a single BEC initially prepared in an optical barrier such as a far-off resonant laser barrier~\cite{pezze05,pezze06}, which splits the single BEC into two separate qualitatively and quantitatively identical pulsed fields. When one of the two pulses is released under gravity, this causes asymmetry of the effective optical potential consisting of the antiharmonic optical trap and the gravitational field, and the two-pulse matter-wave soliton forms two co-propagating unbalanced intensity pulses as a result of gravitational acceleration of one of the two pulses. By measuring the potential energy difference between the two co-propagating pulses, one can, for instance, estimate the kinetic energy due to this gravitational acceleration. In this regard, the analysis of sect. 3 shows that when the coefficient $\mu_1$ of the linear term in the potential is zero, the two pulses are perfectly twinned, indicating the absence of gravitational acceleration of pulses. When $\mu_1$ is nonzero, the two pulses become increasingly unbalanced as $\mu_1$ is increased. Noting that $\mu_1$, according to Eq.~(\ref{eq3c}), is proportional to the constant of gravity $g$ and the mass $m$ of atom species in the condensate, the behavior of the accelerated pulse fraction under the variation of $\mu_1$ reflects an enhancement of the gravitational effect as atoms forming the condensate are heavier.  \\ 
In modeling the double-slit interferometry, an appropriate description of the splitting of a single BEC into two co-propagating matter-wave pulses, taking into account the necessity to accurately control the spatial and temporal locations of each pulse, is essential in probing spatially and temporally varying forces or interactions. While a double-pulse matter-wave field has usually been approximated in variational treatments either by the Hermite\text{--}Gaussian wave or super-sech mode, neither of these two fields is an exact solution to the GP equation governing the system dynamics. \\ 
In this work, we addressed the issue of the consistency of these variational fields versus an exact one-soliton solution to the GP equation obtained by means of an inverse-scattering transform. The following general picture emerges from our analysis:
\begin{itemize}
\item The super-sech mode has a bisoliton shape with respect to the spatial coordinate when a phase factor $\delta$ in its argument is negative. This bisoliton profile becomes sharper, and pulses in the bound state increasingly well seperated, as $\delta$ is decreased in the negative branch. In other words, when $\delta$ is negative but close to zero, the super-sech mode consists of two partially overlapping pulses the separation of which increases as $\delta$ decreases in the negative branch.
\item The Hermite\text{--}Gaussian mode has a bisoliton profile for all nonzero values of its characteristic parameters. However, unlike the super-sech mode, the Hermite\text{--}Gaussian bisoliton is odd with respect to the spatial coordinate. The Hermite\text{--}Gaussian bisoliton describes two pulses that are always partially embedded in each other in the static regime. Therefore, only by propagation can the two pulses deploy themselves into two more or less independent individual entities, as has been observed in numerical simulations in the context of dispersion-managed optical fibers~\cite{Pare99,Pare}. 
\item The NIST soliton solution possesses a permanent bisolitonic profile, in addition, its propagation leads to two individual pulse-shaped entities moving away from each other with different intensities.
\end{itemize}
The results of numerical simulations have emphasized the virtues of the exact soliton solution, and particularly its ability to provide a coherent description of the formation and propagation of two co-existing pulses with distinct dynamical parameters. We therefore conclude that although variational ansatzes such as the Hermite\text{--}Gaussian and super-sech modes are relevant to a qualitative description of atom interferometry involving two-field matter-wave modes, an exact solution to the governing GP equation including a beam splitter and gravity should provide the most physically relevant insight into the system dynamics.  

\begin{acknowledgment} 
The authors thank the referee for drawing their attention to the important work of Satsuma and Yajima (ref. 44) on initial-value problems for the nonlinear Schr\"odinger equation. This work was partially supported by the Academic of Science for the Developing World (TWAS). The authors are grateful to the Physical Society of Japan for financial support in publication.
\end{acknowledgment}

\end{document}